\documentclass[%
nofootinbib,
 amsmath,amssymb,
 aps,
prd,
onecolumn
]{revtex4-2}

\usepackage{graphicx}% Include figure files
\usepackage{graphics}
\usepackage{adjustbox} % Helpful for adjusting table sizes.
\usepackage{dcolumn}% Align table columns on decimal point
\usepackage{bm}% bold math
\usepackage{dsfont}% double-struck symbols
\usepackage{amsthm}% provides proof environment
\usepackage[dvipsnames]{xcolor}
\usepackage{chngcntr}
\usepackage{comment}
\usepackage{apptools}
\AtAppendix{\counterwithin{lemma}{section}}
\usepackage{xy}
\usepackage{qcircuit}
\usepackage{tikz}
\usetikzlibrary{decorations.pathmorphing,decorations.markings}
\usepackage{tkz-euclide}
\usepackage{mathtools}
\usepackage{floatrow} % helps set table text size
\DeclareFloatFont{tiny}{\tiny}% "scriptsize" is defined by floatrow, "tiny" not
\usepackage[capitalise]{cleveref}
\usepackage[T1]{fontenc}
\crefname{figure}{Fig.}{Fig.}

\theoremstyle{definition}

\theoremstyle{definition}

\def\ket#1{\left\vert #1 \right\rangle}
\def\bra#1{\left\langle #1 \right\vert}
\def\braket#1#2{\left\langle #1 \vert #2\right\rangle}

\newcommand{\be}{\begin{equation}}
\newcommand{\ee}{\end{equation}}
\newcommand{\bp}{\begin{pmatrix}}
\newcommand{\ep}{\end{pmatrix}}
\newcommand{\ben}{\begin{enumerate}}
\newcommand{\een}{\end{enumerate}}

\makeatletter
\newcommand{\beq}{%
 \begingroup
 \eqnarray%
 \@ifstar{\nonumber}{}%
}
\newcommand{\eeq}{\endeqnarray\endgroup}
\makeatother

\newcommand{\pd}{\partial}
\newcommand{\es}{& = &}
\newcommand{\rs}{\ = \ }
\newcommand{\nn}{\nonumber\\}

\newcommand{\nm}{\nn & - &}
\newcommand{\nt}{\nn & \times &}

\newcommand{\cG}{\mathcal{G}}
\newcommand{\cH}{\mathcal{H}}
\newcommand{\cL}{\mathcal{L}}
\newcommand{\cM}{\mathcal{M}}
\newcommand{\cN}{\mathcal{N}}
\newcommand{\cP}{\mathcal{P}}
\newcommand{\tdelta}{\tilde\delta}

\usepackage{simpler-wick}
\usepackage{slashed}

\begin{document}

%%TC:ignore
\title{Second-order renormalized Hamiltonian of Yukawa theory}% Force line breaks with \\

\author{Kamil Serafin}
\affiliation{Department of Physics and Astronomy, Tufts University, Medford MA
}%
\author{Carter M. Gustin}
\affiliation{Department of Physics and Astronomy, Tufts University, Medford MA
}%
\author{Peter J. Love}
 \altaffiliation[Also at ]{Brookhaven National Laboratory}
\affiliation{Department of Physics and Astronomy, Tufts University, Medford MA
}%

\begin{abstract}
Using the renormalization group procedure for effective particles
(RGPEP) we calculate the effective Hamiltonians in the theory of
a fermion field coupled to a scalar field via the Yukawa interaction.
The theory is renormalized by the addition of counterterms.
Necessary counterterms are determined by computing matrix
elements of the effective Hamiltonian. All calculations are
performed up to the second order in the expansion in powers
of the coupling constant. Renormalized effective Hamiltonians
are well-defined symmetric forms acting in the Fock space
as opposed to the renormalized bare Hamiltonian, which is
not well-defined without regularization. We introduce
computational techniques that should streamline higher-order
calculations and may be of independent interest.
\end{abstract}

\pacs{Valid PACS appear here} 
\maketitle

\section{Introduction}\label{intro}

Nonperturbative calculations in quantum field theories
such as obtaining the structure of protons and neutrons from
Quantum Chromodynamics are challenging. Currently, the
only ab initio approach to the bound state problems in
QCD that can produce accurate results for some problems
is Lattice QCD. Among those results are mass spectra,
decay constants, magnetic moments and other static
properties of hadrons, as well as studies of fundamental
properties of QCD such as confinement and chiral symmetry
breaking~\cite{Aoki:2013ldr,Davoudi:2022bnl,
ParticleDataGroup:2024cfk,Kamleh:2023gho}.

The success of Lattice QCD would not be possible without
advances in computational power over the last few decades.
Yet, even now, some problems such as deep inelastic
scattering and computation of related structure functions
of a nucleon require additional computational and
algorithmic advances, while others like ab initio
simulation of hadronization seem unfeasible even given
continued advances in high performance
computing~\cite{Cichy:2018mum,Banuls:2019rao,Nagata:2021ugx}.

On the other hand, even near-term quantum computers with
a few hundred qubits will allow Hilbert spaces that are
(many) orders of magnitude greater in dimension than those
representable on the largest classical supercomputers.
While simple qubit counts for classically intractable
problems may not reflect the entirety of the cost of quantum
computing, they incentivize more detailed quantum resource
estimates, and motivate further development of quantum
computers and quantum algorithms. Possible applications where quantum computers
might outperform any classical computers are real-time
simulation of scattering processes in particle collider simulation
from first principles, simulation of strongly coupled
matter at high density or far from equilibrium (relevant
for studies of heavy-ion collisions and studies of neutron
stars), neutrino astrophysics in core-collapse supernovae
and neutron-star mergers, neutrino-nucleus scattering
simulation (relevant for neutrino experiments such as DUNE),
nonequilibrium dynamics of interacting fields for cosmology,
and more~\cite{Bauer:2022hpo,DiMeglio:2023nsa,Jordan:2012xnu,
Martinez:2016yna}.

A particular challenge in simulation of QFT is
renormalization. Canonical or bare Hamiltonians are
typically ill-defined, containing divergences.
Regularization of these bare theories (by cutoffs in
general) is required for any simulation. Renormalization
of the bare parameters and an introduction of counterterms
is necessary to obtain a finite theory, relating
physical inputs (measured masses and coupling constants)
to physical outputs (structure functions, etc).

Renormalization will affect estimates of required resources
for quantum algorithms. Introduction of (finite parts of) the counterterms
will change the structure of the Hamiltonian. Flowing of
coupling constants will change the Hamiltonian's norm.
Changes in cutoffs will change the required number of qubits.
These are all parameters on which quantum resource estimates
for quantum simulation depend.

In this article we develop an approach to relativistic
quantum field theories based on the front form of
Hamiltonian dynamics~\cite{Dirac:1949cp,Brodsky:1997de}
and renormalization group procedure for effective particles
(RGPEP)~\cite{Glazek:2012qj,Glazek:1997sd,Glazek:1993rc,
Glazek:1994qc}. The main, motivating application for the
framework is QCD, but to introduce the new techniques
in a simpler setting we study Yukawa theory
first~\cite{Yukawa:1935xg}. This way we can avoid conceptual
and technical complications arising from gauge symmetry.
Yukawa theory is interesting on its own as a relativistic
model of nuclear interactions~\cite{Machleidt:2022kqp}.
At the same time we provide a complete example of
a relativistic field theory, ready for studies both
on classical high-performance machines and present
and future quantum computers.

In the front form, one considers a hypersurface formed
by a wave front traveling at the speed of light.
Conventionally the wave travels in the direction opposite
to the $z$ axis. Hence, its fronts are described by:
\beq
x^+ \es x^0 + x^3 \rs \text{const.}
\eeq
It is convenient to take the front $x^+ = 0$. This hypersurface
is used to define initial conditions for the field equations.
The field evolution is understood as a succession of states
from one at $x^+$ to another at $x'^+ > x^+$. The evolution
is generated by $P^-/2$, where $P^- = P^0 - P^3$ is the front
form Hamiltonian. For any vector, $v^\pm = v^0 \pm v^3$. The
three coordinates $x^-$, $x^1$, $x^2$ parametrize the
hypersurfaces of fixed front form time $x^+$. The particle's
$p^-$ is its front-form energy, while $p^+$ and $p^\perp
= (p^1, p^2)$ are its longitudinal and transverse momenta,
respectively. The condition, $p_\mu p^\mu = m^2$ implies
the dispersion relation,
\beq
p^- \es \frac{m^2 + (p^\perp)^2}{p^+}
\ .
\label{eq:dispersion}
\eeq

An important feature of the front form is the fact that
for any massive particle its longitudinal momentum is
strictly positive. Using the usual relations for momenta,
$p^+ = \sqrt{m^2 + (p^\perp)^2 + (p^3)^2} + p^3$, hence,
even for very large negative $p^3$, $p^+$ will always remain
positive.

The front form of Hamiltonian dynamics presents a unique
set of advantages. Firstly, the vacuum state in the front
form is trivial, i.e., it is the same in the free and the
interacting version of the theory. We ensure that this is
the case by introducing a cutoff on small $p^+$ momentum
modes and the resulting Hamiltonian is called the cutoff
Hamiltonian. The cutoff Hamiltonian may need to be
supplemented with special counterterms that reproduce the
vacuum effects removed by the cutoff~\cite{Wilson:1994fk}.
Such counterterms are not expected in the Yukawa theory.
Triviality of the vacuum means that the problem of particle
bound states does not require one to solve for a complicated
vacuum state first, as is the case in instant time
approaches. In the context of quantum computing, this
means that there is no need for costly preparation
of a highly entangled vacuum state.

Another advantage of the front form is boost invariance
of the wave functions -- the internal structure of the
hadron is described by the same relative-motion wave
function regardless of the motion of the hadron as
a whole~\cite{Brodsky:1997de}. This fact is highly
advantageous in the context of computing hadron structure
functions in processes such as deep inelastic scattering.
A simple demonstration is given by the formula for the
invariant mass of two particles: $\cM_{12}^2 = (p_1 + p_2)^2
= (m_1^2 + k^2)/x + (m_2^2 + k^2)/(1-x)$, where
$x = p_1^+/(p_1^+ + p_2^+)$, and $k = (1-x) p_1^\perp
- x p_2^\perp$ are relative longitudinal and transverse
momenta, respectively. The invariant mass of two particles
depends only on relative momenta, $x$ and $k$ that are
invariant under Lorentz transformations that preserve
the $x^+ = 0$ hypersurface. 

Any approach to relativistic QFTs has to address
the problem of renormalization. The canonical
Hamiltonian of Yukawa theory is not well-defined
without regularizing the interactions, and the
observables diverge when the regularization
is gradually removed. In Hamiltonian approaches,
the front form in particular, symmetries are often
not explicitly conserved, hence, regularization
typically breaks more symmetries than in Lagrangian
approaches and hence Hamiltonians suffer more severe
singularities. These difficulties lead to the
development of powerful renormalization techniques.
In this work we adopt the renormalization group
procedure for effective particles (RGPEP)~\cite{Glazek:2012qj},
which is an implementation of the similarity
renormalization group~\cite{Glazek:1993rc,Glazek:1994qc}.

RGPEP deals with ultraviolet divergences of local QFTs
by defining effective particles whose interactions are
no longer local and by expressing the initial Hamiltonian
in terms of the effective particles. Nonlocal interactions
make the effective Hamiltonians energetically narrow,
i.e., in the basis of states with increasing energy
expectation values the Hamiltonian matrix is approximately
band diagonal. The band limits the momenta allowed in
the loop integrals, which can then no longer produce divergences.
The divergent nature of the initial Hamiltonian is exhibited
in diverging matrix elements of the effective Hamiltonians.
Therefore, one can identify the divergent terms and
the form of the counterterms that are necessary to remove
the divergences.

An important implication of the narrowness of the
effective Hamiltonians is that once the counterterms
are included, one can remove the regularization and
the effective Hamiltonians remain well-defined
operators in the Fock space.\footnote{For simplicity
of formal manipulations we use word ``operator'' to
describe both actual operators and quadratic forms
that need not correspond to operators. The distinction
is not relevant for the discussion until
Sec.~\ref{sec:renoYukawa} where we do make mathematically
precise statements about the Hamiltonian.}
This is not the case for the canonical Hamiltonian
supplemented with appropriate counterterms -- the
coefficients in the Hamiltonian, hence also
the matrix elements of the Hamiltonian, diverge
when the regularization is removed.

There is one drawback of the renormalization procedure
we adopt that needs to be mentioned. In the effective
Hamiltonians the front-form longitudinal boost invariance
is no longer conserved exactly. This is a consequence
of the choice of the generator in the RGPEP equation.
There exist generators that do not lead to this
problem~\cite{Glazek:2012qj} which could be used in Yukawa
theory. Our choice is motivated by advantages one gains
when the approach is applied to QCD, i.e., infrared
divergences cancel in the color-singlet subspace of
the Fock space~\cite{Serafin:2023pkf}. In a nonperturbative
calculation any divergences that survive can invalidate
the calculation. Therefore, one needs to prioritize the
cancellation of divergences over boost invariance.
In practice, boost-invariance breaking means that the masses
of self-bound systems depend on the total longitudinal
momenta of these systems. The longitudinal boost invariance
is directly linked with the invariance with respect to
the scale parameter of RGPEP. Finding solutions to the RGPEP
equation requires in general some level of approximation
because exact solutions can only be found in simplified
scenarios~\cite{Glazek:2021vnw,Glazek:2012am,Glazek:2013gba,
Maslowski:2024gwt}. Therefore, any approximation made while
calculating the effective Hamiltonians results in an
approximation to boost invariance. This drawback does not
necessarily mean that one entirely loses the advantage
of the front form approach because one can systematically
improve the precision of the calculation of the effective
Hamiltonians. We provide the first RGPEP calculation for
Yukawa theory which can be systematically improved in
future work.

In the approach presented in this paper, any theory is
solved in two steps: renormalization, and diagonalization
of the Hamiltonian. The renormalization step is performed
using perturbative expansion in powers of the coupling
constant. The effective Hamiltonians include terms
of all orders in the coupling constant, but we only
keep terms up to the second order. The diagonalization
step can be performed using a numerical procedure
of one's choice. Popular choices for front-form
Hamiltonians include discretized light-cone quantization
(DLCQ)~\cite{Pauli:1985pv,Pauli:1985ps,Eller:1986nt,
Harindranath:1987db,Hornbostel:1988fb,Hornbostel:1988ne,
Vary:2021cbh} and basis light-front quantization
(BLFQ)~\cite{Vary:2009gt,Wiecki:2014ola,Li:2017mlw,
Jia:2018ary,Lan:2019vui,Hu:2020arv,Lan:2021wok,
Xu:2021wwj,Nair:2022evk,Kuang:2022vdy,Lin:2024ijo}.
As a Hamiltonian approach, the front form is one of
the natural candidates for simulations on future
quantum computers~\cite{Kreshchuk:2020dla,
Kreshchuk:2020kcz,Kreshchuk:2020aiq,Kirby:2021ajp,
Kreshchuk:2023btr,Qian:2024gph,Simon:2025pbo}.

Our calculation is also one of a few in which the full
renormalized Hamiltonian of a quantum field theory is
calculated up to second order in the expansion in the
coupling constant. The simplest nontrivial quantum field
theory of a scalar field $\phi$ with $\phi^3$ interactions
has been used in the past as a stepping stone for
calculations in pure-glue quantum
chromodynamics~\cite{Allen:1998bk,Kylin:1998wr,Allen:1999kx}.
In 1+5 dimensions $\phi^3$ theory exhibits asymptotic freedom.
Therefore, the three-point interaction $\phi^3$ in 1+5D
closely resembles the analogous three-point interaction
in QCD. The simplicity of $\phi^3$ theory allows one to
easily write down the complete effective Hamiltonian up
to second order in the coupling constant. The same theory,
but in 1+3D, has been used for exemplary calculation of
the full interacting Poincare algebra in the front form
of Hamiltonian dynamics up to second order, including
the effective Hamiltonian and the other two dynamical
generators~\cite{Glazek:2001uw}.

Apart from the final result, our calculational techniques
are also of independent interest. Even though the ideas
and techniques we use, such as Wick's theorem, are well-known,
they have not been employed in the context of RGPEP. More
powerful techniques were required because the renormalization
flow equation generates complicated expressions that can
obscure the possibility to come up with simple formulas.
Our techniques simplify the calculations and enable us to
write down short formulas that describe many different
interaction vertices all at once.

The remainder of the paper is organized as follows.
In Sec.~\ref{sec:effectiveParticles} the RGPEP is
introduced. Section~\ref{sec:reno} describes the
strategy of finding counterterms that ensure effective
Hamiltonians are well-defined. Section~\ref{sec:yukawa}
presents the main results and is divided into subsections.
The canonical Hamiltonian of Yukawa theory, defined
in Sec.~\ref{sec:canonical}, is regularized in
Sec.~\ref{sec:regularization}. In Sec.~\ref{sec:effHam}
we begin calculating the effective Hamiltonians.
Section~\ref{sec:wick} introduces the Wick's
diagrams and the Wick's theorems that are used in
Secs.~\ref{sec:H2tree} and \ref{sec:H2self} to evaluate
contributions to the effective Hamiltonians from tree
diagrams and diagrams containing loops, respectively.
In Sec.~\ref{sec:renoYukawa} we compute matrix elements
of the effective Hamiltonians and determine the
counterterms. Section~\ref{sec:summary} summarizes
all contributions to the renormalized Hamiltonians.
We conclude the article in Sec.~\ref{sec:conclusion}
while some notation and conventions are described
in Appendix~\ref{sec:notation}.

\section{Effective particles}
\label{sec:effectiveParticles}

In this section we introduce the main definitions and
concepts needed to perform calculations in the framework
of RGPEP.

Suppose that $q(i)$ is an annihilation operator for a
particle characterized by $i$ which represents all
relevant quantum numbers, e.g., momentum, front-form
helicity, color, as well as the type of the particle
such as quark (including its flavor), scalar boson, etc.
$q(i)^\dagger$ is the corresponding creation operator.
We define effective operators,
\beq
q_t(i)
\es
U_t \, q(i) \, U_t^\dag
\ ,
\eeq
where $U_t$ is a unitary operator. Both $U_t$ and the effective
particles are parametrized by $t \ge 0$. The initial
operators are defined at $t = 0$. Hence, $U_0 = 1$, and
$q_0 = q$. In relativistic quantum field theories $q(i)$
correspond to pointlike, bare particles. For $t > 0$,
our construction will give $q_t(i)$ which correspond to
effective particles that interact
nonlocally~\cite{Glazek:2010zr}. Hence, one can think of
them as having some finite size. The larger the $t$, the
larger the size of the effective particles.

Using bare particle operators one can build the bare Fock
space. Using effective particle operators one can build
the effective Fock space. As long as the unitary operator
$U_t$ is well-defined, the two Fock spaces are equivalent,
with states $|\psi\rangle_t$ in the effective Fock space,
expressible in terms of the states $|\psi\rangle$ in the
bare Fock space,
\beq
|\psi\rangle_t
\es
U_t |\psi\rangle .
\eeq
The change from bare to effective particles, therefore,
can be regarded as a change of basis.

The effective Hamiltonian, $H_t$ is defined by the following
set of equations:
\beq
H_t \es H_0 \rs \cH_0 \ ,
\\
\frac{d}{dt} \cH_t
\es
\left[ \cG_t, \cH_t \right]
 ,
\label{eq:RGPEP}
\eeq
where
\beq
\cH_t
\es
U_t^\dag H_t \, U_t
\ ,
\eeq
and $\cG_t = \left[ \cH_f, \cH_t \right]$ is the generator
of infinitesimal unitary transformation with $\cH_f$ being
the free Hamiltonian. Those equations fix the unitary operator
to be the solution of the following set of equations,
\beq
U_t^\dag \, \frac{d}{dt} U_t
\es
- \cG_t
 ,
\label{eq:Ut}
\\
U_0 \es 1 \ .
\eeq

Different choices of the generator are
possible~\cite{Glazek:1994qc,Mielke:1998ag,Glazek:2002iw,
Glazek:2003xr,Wegner2006,Glazek:2008pg,Anderson:2008mu,
Glazek:2012qj}. In most cases the goal is to diagonalize
the Hamiltonian, but one might want to block-diagonalize
it instead, see Ref.~\cite{Anderson:2008mu}. In numerical
calculations it is advantageous to use a softer generator,
i.e., one that does not produce too abrupt changes of the
matrix elements of the Hamiltonian, because then the step
size can be larger and is more predictable. Softer
generators can also greatly improve the rate of convergence
of perturbative calculations of the effective
Hamiltonians~\cite{Glazek:2002iw,Glazek:2003xr}.
One can also design a generator that does not lead to
the breaking of the kinematic Lorentz symmetries of the
front form~\cite{Glazek:2012qj}. Our choice gives a rather
simple form of the RGPEP equations, which is desirable
at this stage of the development. However, most importantly,
it allows for cancellation of infrared divergences in
the color-singlet sector in QCD~\cite{Serafin:2023pkf}.
As a consequence, the exact longitudinal boost invariance
is lost, but nonperturbative diagonalization of the
Hamiltonian in truncated color-singlet subspaces is
well-defined, which is a prerequisite for numerical studies.
Further refinements to the generator, such as those in
Refs.~\cite{Glazek:2002iw,Glazek:2003xr}, are best studied
once the numerical tools for solving Hamiltonians of
QFTs are better developed.

The difference between $H_t$, $H_0$, and $\cH_t$ can be
illustrated in general with the following equations (the
details of notation are defined in App.~\ref{sec:notation}):
\beq
H_t
\es
\sum_{n = 2}^\infty
\sum_{i_1, \dots, i_n}
c_t(i_1, \dots, i_n)
\ q_t(i_1)^\dag \dots q_t(i_n)
\ .
\eeq
$H_t$ is the effective Hamiltonian which is written
in terms of effective creation and annihilation
operators $q_t$ and effective coefficients $c_t$.
\beq
H_0
\es
\sum_{n = 2}^\infty
\sum_{i_1, \dots, i_n}
c_0(i_1, \dots, i_n)
\ q(i_1)^\dag \dots q(i_n)
\ .
\eeq
$H_0$ is the initial, bare Hamiltonian written in terms of
bare operators $q$ and bare coefficients $c_0$. The effective
Hamiltonian and the bare one represent the same abstract
operator, just written using different bases defined by
operators $q_t$ and $q$, respectively. Hence, $H_t = H_0$.
\beq
\cH_t
\es
\sum_{n = 2}^\infty
\sum_{i_1, \dots, i_n}
c_t(i_1, \dots, i_n)
\ q(i_1)^\dag \dots q(i_n)
\ .
\eeq
$\cH_t$ is written in terms of bare operators, but with effective
coefficients. This way the derivative with respect to $t$ applied
to $\cH_t$ acts on the coefficients only, while applied to $H_t$
acts also on the effective operators, which depend on $t$ in such
a way that $dH_t/dt = 0$.

We assume that the free Hamiltonian has the form,
\beq
\cH_f
\es
\sum_i \, p_i^- \, q(i)^\dag q(i)
\ ,
\eeq
where $p_i^- = [m_i^2 + (p_i^\perp)^2]/p_i^+$ is the front-form
energy of the particle characterized by $i$. Using,
$\cH_f q(i)^\dagger = q(i)^\dagger \cH_f + p_i^- q(i)^\dagger$
and $\cH_f q(i) = q(i) \cH_f - p_i^- q(i)$, it is easy to show
that
\beq
\left[ \cH_f, \cH_t \right]
\es
\sum_{n = 2}^\infty
\sum_{i_1, \dots, i_n}
\left( P_a^- - P_b^- \right)
c_t(i_1, \dots, i_n)
\ q(i_1)^\dag \dots q(i_k)^\dag
  q(i_{k+1}) \dots q(i_n)
\ ,
\eeq
where $P_a^-$ is the sum of $p^-$ over the particles that
are created in the interaction, while $P_b^-$ is the sum
of $p^-$ over the particles that are annihilated in the
interaction. For example, if there are $n$ operators,
the first $k$ of them are creation operators, and the last
$n-k$ of them are annihilation operators, then
\beq
P_a^-
\es
\sum_{r = 1}^k \, p_{i_r}^-
\ ,
\\
P_b^-
\es
\sum_{r = k + 1}^n \, p_{i_r}^-
\ .
\eeq
Thus, the generator closely resembles the Hamiltonian $\cH_t$.
The only difference is that the coefficients $c_t$ in $\cH_t$
are replaced with $(P_a^- - P_b^-) c_t$ in the generator. In
fact, the commutator $\left[ \cH_f, \cH_t \right]$ defines a
linear operator acting on $\cH_t$. We define,
\beq
\cG_t
\es
\left[ \cH_f , \cH_t \right]
\rs
-i\pd_f^- \cH_t
\ ,
\eeq
where $\pd_f^-$ means ``take the front-form time derivative
$\pd^- = 2 \frac{\pd}{\pd x^+}$, as if the operators evolved
according to free evolution equations, i.e., $q_t(i, x^+)
= e^{-i p_i^- x^+/2} q_t(i)$.'' We could drop the subscript
``$f$'' if we were working in the interaction picture. The
actual time evolution of the creation and annihilation
operators in the Heisenberg picture is not known in general.
The operator $i\pd_f^-$ can be understood as counting the
difference between kinetic energy (as defined by $\cH_f$)
before and after the interaction.

Equation~(\ref{eq:RGPEP}) is in general very difficult to
solve. For QCD, due to asymptotic freedom, a perturbative
expansion in powers of the coupling constant should be
a viable approach. For Yukawa theory, we assume that the
coupling constant at the scales we are considering is small
enough to use the perturbative expansion. We define the
interacting Hamiltonian,
\beq
\cH_{It} \es \cH_t - \cH_f \ .
\eeq
Therefore,
\beq
\frac{d}{dt} \cH_t
\es
-(i\pd_f^-)^2 \cH_{It}
+ \left[ -i\pd_f^- \cH_{It}, \cH_{It} \right]
 .
\label{eq:RGPEP2}
\eeq
where we used the fact that $\pd_f^- \cH_f = 0$ to replace
$\pd_f^- \cH_t$ with $\pd_f^- \cH_{It}$. An immediate implication
is $d\cH_f/dt = 0$. The second term on the rhs of
Eq.~(\ref{eq:RGPEP2}) is of higher order than the first term.
Therefore, the leading-order solution that neglects the
second term is $e^{-t(i\pd_f^-)^2} \cH_{I0}$. This suggests
the following form of a general solution,
\beq
\cH_{It}
\es
e^{-t(i\pd_f^-)^2} h_t
\ ,
\label{eq:ht}
\eeq
where for convenience $h_t$ is called the reduced Hamiltonian.
The RGPEP equation becomes,
\beq
\frac{d}{dt} h_t
\es
e^{t(i\pd_f^-)^2}
\left[ -i\pd_f^- \cH_{It}, \cH_{It} \right] .
\eeq
We assume that the interaction Hamiltonian, and therefore,
the reduced Hamiltonian admit an expansion in powers of
a single coupling constant $g$,
\beq
h_t \es h_{t , 1} + h_{t , 2} + h_{t , 3} + \dots \ ,
\eeq
where $h_{t,n}$ is of order $g^n$, or more precisely,
$h_{t,n}/g^n$ is independent of $g$. We arrive at a set
of equations, order by order,
\beq
\frac{d h_{t , 1}}{dt} \es 0 \ ,
\label{eq:RGPEP1st}
\\
\frac{d h_{t , 2}}{dt}
\es
e^{t(i\pd_f^-)^2}
\left[ -i\pd_f^- \cH_{t,1}, \cH_{t,1} \right]
 ,
\label{eq:RGPEP2nd}
\\
\frac{d h_{t , 3}}{dt}
\es
e^{t(i\pd_f^-)^2}
\left[ -i\pd_f^- \cH_{t,1}, \cH_{t,2} \right]
+
e^{t(i\pd_f^-)^2}
\left[ -i\pd_f^- \cH_{t,2}, \cH_{t,1} \right]
 ,
\quad\text{etc.,}
\label{eq:RGPEP3rd}
\eeq
where $\cH_{t,k} = e^{-t(i\pd_f^-)^2} h_{t,k}$. At any order
$k$ the derivative $dh_{t,k}/dt$ is expressed in terms of
reduced Hamiltonians in all lower orders $h_{t,1}$, \dots,
$h_{t,k-1}$, hence the perturbation theory is established.
Further calculations are best illustrated using an example;
see Sec.~\ref{sec:yukawa}.

\section{Renormalization in perturbation theory}
\label{sec:reno}

In this section we summarize the principles of renormalization
of Hamiltonians~\cite{Glazek:1993rc}.
Most quantum field theories of interest, Yukawa theory
included, lead to divergent results when one attempts
to compute observables such as the energy spectrum of the
Hamiltonian. The source of those divergences can be traced
to the local character of the interactions. With local
interactions, matrix elements of the Hamiltonian between
states of arbitrarily different energies can be not only
nonzero but large. This implies that phenomena at vastly
different energy scales are strongly coupled.

The goal of the RGPEP evolution is to force the effective
Hamiltonian, $\cH_t$ into a band diagonal form, thereby
decoupling particles at scales that differ by more than
the width of the band. By band-diagonal form, we mean that
matrix elements that are sufficiently far from the diagonal
(according to some metric) are required to be zero or
suppressed to the degree that they may be neglected.
This property of the effective Hamiltonians is called
narrowness.

The generator we chose does ensure that the effective
Hamiltonians are narrow. According to Eq.~(\ref{eq:ht}),
if we consider eigenstates of the free Hamiltonian, say
$\ket{\psi_j}$ and $\ket{\psi_k}$, such that $\cH_f
\ket{\psi_j} = P_j^- \ket{\psi_j}$ and $\cH_f \ket{\psi_k}
= P_k^- \ket{\psi_k}$, then the matrix element of the
interaction Hamiltonian $\bra{\psi_j} \cH_{It} \ket{\psi_k}
= e^{-t(P_j^- - P_k^-)^2} \bra{\psi_j} h_t \ket{\psi_k}$.
As long as the reduced Hamiltonian does not develop very
large matrix elements, the matrix elements of the effective
Hamiltonian become very small -- irrelevant in practice --
when the difference $|P_j^- - P_k^-|$ is considerably larger
than $1/\sqrt{t}$. The exponential factor,
$e^{-t(P_j^- - P_k^-)^2}$ is called the RGPEP form factor,
or simply, the form factor.

Form factors, as the name suggests, make the effective
interactions nonlocal and ensure that no divergences can
appear in calculations where they would normally appear
in local field theories. However, the effective theory
with nonlocal interactions is exactly equivalent to the
initial theory with local interactions as long as the
initial theory is regularized first. The divergent
nature of the initial theory is manifested in diverging
matrix elements of the effective Hamiltonian. Therefore,
the goal of renormalization is to make the effective
Hamiltonian well-defined in the limit of the regularization
being lifted. In effect, one has to modify the initial
Hamiltonian, but the same limit of the modified initial
Hamiltonian need not be well-defined. For example,
the bare coupling can approach infinity as in the Wilson
model~\cite{Wilson:1970tp}.

The renormalization program in perturbation theory is
as follows. Firstly, one has to define the initial
Hamiltonian (for example, the canonical Hamiltonian)
and regularize it. Regularization introduces a cutoff
parameter and the limit as this parameter goes to zero
(alternatively to infinity) corresponds to lifting the
regularization. The regularization needs to remove all
divergences in the theory, since, otherwise,
the unregulated expression for the Hamiltonian lacks
the mathematical meaning necessary for computations.
Secondly, one calculates the effective Hamiltonian
as a solution to the RGPEP Eq.~(\ref{eq:RGPEP}).
One can do this order by order in perturbation theory,
see Eqs.~(\ref{eq:RGPEP1st})--(\ref{eq:RGPEP3rd}). In
perturbation theory, in order to compute the $n$th order
Hamiltonian, renormalization in orders $1$ through $n-1$
needs to be completed first. Assuming this is the case,
at the $n$th order one needs to ensure that all matrix
elements between states from the domain of the effective
Hamiltonian are finite when the cutoff parameter is sent
to zero. If any matrix element diverges when the cutoff
is sent to zero, then one needs to supplement the initial
Hamiltonian with appropriate counterterms. At the $n$th
order of the calculation, the counterterms are of order
$g^n$ and can be determined in the following way.
Equations~(\ref{eq:RGPEP1st})--(\ref{eq:RGPEP3rd})
are integrated from $t=0$ to $t$. The left hand side
after integration contains $h_{t,n} - h_{0,n}$, where
$h_{t,n}$ is the effective reduced Hamitonian in order
$n$, and $h_{0,n}$ is the initial condition, bare
Hamiltonian in order $n$. The right hand side is fixed
by lower-order results. If it produces a divergent term,
then the divergence needs to be isolated and inserted
with an opposite sign into the initial condition term,
$h_{0,n}$. Therefore, the initial condition plus the
term that is produced by RGPEP evolution in order $g^n$
sum to a finite expression.

The renormalization process does not end with the removal
of divergences because counterterms are not uniquely
determined. Two counterterms that differ by only a constant
remove the divergence equally well. Therefore, one needs to
determine not only the infinite parts of the counterterms,
but also their finite parts. Symmetries and coupling coherence
can be used to constrain the finite parts. For example,
the Lorentz covariance of the scattering matrix introduces
such constraints~\cite{Maslowski:1997mn,Serafin:2017vld}.
The main idea of coupling coherence is that only a finite
number of parameters are independent functions of the
renormalization parameter. All other parameters and
effective interactions, be them infinitely many and however
complicated, are uniquely determined by the independent
parameters~\cite{Perry:1993gp,Perry:2001je}. The final step,
once all theoretical arguments have been exhausted,
is computation of observables and comparison with
experiment to fix any remaining free parameters.

\section{Yukawa theory}
\label{sec:yukawa}

We present the step by step process of calculating the
renormalized Hamiltonian of the Yukawa field theory.
We start with the canonical Hamiltonian, regularize it,
compute the effective Hamiltonian, and end with the computation
of the matrix elements and determination of the counterterms.
On the way we develop Wick diagrams as a useful tool
to represent, organize, and manipulate complicated
interaction terms of the Hamiltonian of a quantum
field theory.

\subsection{Canonical Hamiltonian of Yukawa theory}
\label{sec:canonical}

We start from the Lagrangian:
\beq
\cL
\es
  \bar\psi \left(i\slashed\pd - m \right) \psi
+ \frac{1}{2} \pd_\mu \phi \, \pd^\mu \phi
- \frac{1}{2} \mu^2 \phi^2
- g \bar\psi \psi \phi
\ ,
\eeq
where $\phi$ is a scalar field, $\psi$ is a fermion field,
$\mu$ and $m$ are boson and fermion masses, respectively,
and $g$ is a dimensionless coupling constant. The canonical
Hamiltonian is the integral of the Hamiltonian density,
$\cH$ over the hypersurface defined by $x^+ = 0$,
\beq
H_\text{canonical}
\es
\int dx^- d^2 x^\perp \, \cH
\ .
\eeq
The canonical Hamiltonian density in the front form of
Hamiltonian dynamics is $\cH = \frac{1}{2} T^{+-}$, where
$T^{\mu\nu}$ is the canonical stress-energy tensor
obtained as the conserved Noether's current due to
invariance with respect to translations in spacetime.
For Yukawa theory,
\beq
\cH
\es
  \cH_{\psi^2 + \phi^2}
+ \cH_{\psi^2\phi}
+ \cH_{\psi^2 \phi^2}
\ ,
\eeq
where
\beq
\cH_{\psi^2 + \phi^2}
\es
  \cN\!\left(
    \bar\psi \frac{\gamma^+}{2}
    \frac{ (i\pd^\perp)^2 + m^2 }{i\pd^+} \psi
  \right)
+ \cN\!\left\{
    \frac{1}{2} \phi \left[ (i\pd^\perp)^2 + \mu^2 \right] \phi
  \right\}
 ,
\\
\cH_{\psi^2 \phi}
\es
g \cN\!\left(\bar\psi \psi \phi\right) ,
\\
\cH_{\psi^2 \phi^2}
\es
\frac{1}{2} g^2
\cN\!\left(\bar\psi \phi \frac{\gamma^+}{i\pd^+} \phi \psi \right) .
\eeq
$\cN$ stands for normal ordering, i.e., sorting factors of
an expression until all creation operators are to the left
of all the annihilation operators and multiplying the
expression by minus sign whenever two fermionic operators
are transposed. The fields are quantized on the spacetime
hyperplane parametrized by $x^\mu = (x^+ = 0, x^-, x^\perp)$,
and are expanded in a plane-wave basis,
\beq
\psi(x)
\es
\sum_{\sigma} \int\frac{dp^+ d^2p^\perp}{16\pi^3 p^+} \theta(p^+)
\left[
  u_\sigma(p) e^{-i p x} b_{p \sigma}
+ v_\sigma(p) e^{ i p x} d_{p \sigma}^\dag
\right]
 ,
\\
\phi(x)
\es
\int\frac{dp^+ d^2p^\perp}{16\pi^3 p^+} \theta(p^+)
\left(
  e^{-i p x} a_{p}
+ e^{i p x} a_{p}^\dag
\right)
 ,
\eeq
where $b$ and $d$ are fermionic annihilation operators,
$a$ is a bosonic annihilation operator, $u$ and $v$ are
spinors for particles and antiparticles, $p = (p^+, p^\perp)$
is the light-form momentum, $\sigma$ is the light-front
helicity, and $\theta$ is the Heaviside theta function.
In this work we do not consider time evolution and we keep
$x^+ = 0$. More generally, the phase factors in the interaction
picture are $p x = p^- x^+/2 + p^+ x^-/2 - p^\perp x^\perp$,
where $p^-$ is defined in Eq.~(\ref{eq:dispersion}) with
the appropriate mass, $m$ for the fermionic field and $\mu$
for the bosonic field.

For our purposes it is convenient to work in the momentum-space
representation,
\beq
\Phi(q)
\es
\int dx^- d^2 x^\perp
\ e^{\frac{i}{2} q^+ x^- - i q^\perp x^\perp}
\ \phi(x)
\\
\es
\frac{
  \theta(q^+) a_{q}
+ \theta(-q^+) a_{-q}^\dag
}{|q^+|}
\ ,
\\
\Psi(q)
\es
\int dx^- d^2 x^\perp
\ e^{\frac{i}{2} q^+ x^- - i q^\perp x^\perp}
\ \psi(x)
\\
\es
\sum_{\sigma}
\frac{
  \theta(q^+) u_{\sigma}(q) b_{q \sigma}
+ \theta(-q^+) v_{\sigma}(-q) d_{-q \sigma}^\dag
}{|q^+|}
\ .
\eeq
Importantly, $q^+$ can be both positive and negative, and
its sign determines whether the field contains the creation
or the annihilation particle operator. This is to be contrasted
with the instant form in which fields are split into, and
always contain, both positive and negative frequency components.
In the front form one can identify positive frequency solutions
with $q^+ > 0$ and negative frequency solutions with $q^+ < 0$.
Note also that $\Phi$ and $\Psi$ are functions of $q^+$ and
$q^\perp$ only (apart from $x^+$ which is always fixed to zero).
One can assign $q^-$ components in accordance to free evolution,
\beq
i\pd_f^- \Phi(q)
\es
\left. q^- \right|_\mu \Phi(q)
\ ,
\\
i\pd_f^- \Psi(q)
\es
\left. q^- \right|_m \Psi(q)
\ ,
\\
i\pd_f^- \bar\Psi(-q)
\es
\left. q^- \right|_m \bar\Psi(-q)
\ ,
\eeq
where the minus component is annotated, i.e.,
$\left. q^- \right|_m = \frac{m^2 + (q^\perp)^2}{q^+}$,
and $\left. q^- \right|_\mu = \frac{\mu^2 + (q^\perp)^2}{q^+}$.

In the momentum space representation the Hamiltonian is,
\beq
H_\text{canonical}
\es
  H_{\psi^2 + \phi^2}
+ H_{\psi^2\phi}
+ H_{\phi^3}
+ H_{\psi^2 \phi^2}
\ ,
\eeq
where
\beq
H_{\psi^2 + \phi^2}
\es
\int[q] \left. q^- \right|_m
\,\cN\!\left[
    \bar\Psi(q)
    \frac{\gamma^+}{2}
    \Psi(q)
  \right]
+
\int[q] \, \frac{q^+ \left. q^- \right|_\mu}{2}
\,\cN\!\left[ \Phi(-q) \Phi(q) \right] ,
\label{eq:yukawaFreeHamiltonian}
\\
H_{\psi^2\phi}
\es
g
\int[q_1 q_2 q_3]
\,\tdelta_{123}
\ \cN\!\left[ \bar\Psi(-q_1) \Psi(q_2) \Phi(q_3) \right] ,
\label{eq:Hpsi2phi}
\\
H_{\psi^2 \phi^2}
\es
g^2
\int[q_1 q_2 q_3 q_4]
\,\tdelta_{1234}
\ \cN\!\left[
\bar\Psi(-q_1) \Phi(q_3)
\frac{\gamma^+/2}{q_2^+ + q_4^+}
\Phi(q_4) \Psi(q_2)
\right] .
\label{eq:Hpsi2phi2}
\eeq
where
\beq
\tdelta_{1\dots n}
\rs
16\pi^3 \delta(q_1^+ + \dots + q_n^+)
\delta^2(q_1^\perp + \dots + q_n^\perp)
\ ,
\\
\left[q\right] \rs \frac{dq^+ d^2 q^\perp}{16\pi^3} \ ,
\quad
[q_1 q_2] \rs [q_1] [q_2] \ ,
\quad\text{etc.}
\eeq
The mass in the minus component $q^-$ can usually be
inferred easily from the label. For example, $q_3$ in
Eq.~(\ref{eq:Hpsi2phi}) corresponds to the scalar field,
therefore, $q_3^-$ is defined with mass $\mu$, while $q_1^-$
and $q_2^-$ are defined with mass $m$, because they correspond
to the fermion fields. Henceforth, for convenience we drop
the mass annotations.

It is worth noting that the interaction terms are written
in a very concise way. All $q_i^+$ are integrated from
$-\infty$ to $\infty$ and are equal to the actual momenta
only up to sign. For example, $H_{\psi^2\phi}$, depending
on the signs of the momentum variables $q_1^+$, $q_2^+$,
and $q_3^+$, will give eight different types of interactions,
see Table~\ref{tab:3list}. For $q_1^+ > 0$, $\bar\Psi(-q_1)$
is proportional to $d_{q_1}$ and the actual momentum is $p_1
= q_1$. For $q_1^+ < 0$, $\bar\Psi(-q_1)$ is proportional
to $b_{-q_1}^\dagger$ and the actual momentum is $p_1 = -q_1$.
Similarly, the sign of $q_2^+$ is related to what $\Psi(q_2)$
contributes in Eq.~(\ref{eq:Hpsi2phi}), and the sign of
$q_3^+$ determines the contribution from $\Phi(q_3)$.
Equations~(\ref{eq:Hpsi2phi}) and (\ref{eq:Hpsi2phi2}) are
written in a way that ensures $p_i^\mu = \mathrm{sgn}(q_i^+)
q_i^\mu$. Note that since the sign of $q^+$ determines the
sign of $q^-$, $p_i^- = \mathrm{sgn}(q_i^+) q_i^-$ also holds.
Combinations $+++$ and $---$ correspond to annihilation of
particles to vacuum and creation of particles from the vacuum,
respectively, and are absent in the cutoff Hamiltonian.
Therefore, $H_{\psi^2 \phi}$ has six distinct interaction
terms. $H_{\psi^2 \phi^2}$ has ten distinct interaction terms
(four combinations from $\bar\Psi(-q_1) \Psi(q_4)$ times
three combinations from $\Phi(q_2) \Phi(q_3)$\footnote{$q_2^+,
-q_3^+ > 0$, and $-q_2^+, q_3^+ > 0$ cases are considered
of the same type for $\Phi(q_2) \Phi(q_3)$.} minus two
zero-mode terms). Similar notation has been introduced
before~\cite{Glazek:2010zr}.

\begin{table}[h]
\begin{tabular}{c|c|ccc}
$q_1^+ q_2^+ q_3^+$ &
Interaction & $p_1$ & $p_2$ & $p_3$ \\
\hline
$+++$ & $d_{q_1} b_{q_2} a_{q_3}$ & $q_1$ & $q_2$ & $q_3$ \\
$++-$ & $d_{q_1} b_{q_2} a_{-q_3}^\dagger$ & $q_1$ & $q_2$ & $-q_3$ \\
$+-+$ & $d_{q_1} d_{-q_2}^\dagger a_{q_3}$ & $q_1$ & $-q_2$ & $q_3$ \\
$+--$ & $d_{q_1} d_{-q_2}^\dagger a_{-q_3}^\dagger$ & $q_1$ & $-q_2$ & $-q_3$ \\
$-++$ & $b_{-q_1}^\dagger b_{q_2} a_{q_3}$ & $-q_1$ & $q_2$ & $q_3$ \\
$-+-$ & $b_{-q_1}^\dagger b_{q_2} a_{-q_3}^\dagger$ & $-q_1$ & $q_2$ & $-q_3$ \\
$--+$ & $b_{-q_1}^\dagger d_{-q_2}^\dagger a_{q_3}$ & $-q_1$ & $-q_2$ & $q_3$ \\
$---$ & $b_{-q_1}^\dagger d_{-q_2}^\dagger a_{-q_3}^\dagger$ & $-q_1$ & $-q_2$ & $-q_3$ \\
\end{tabular}
\caption{\label{tab:3list}Type of the interaction term
in $H_{\psi^2\phi}$ depending on the signs of $q_1^+$,
$q_2^+$, and $q_3^+$ and the actual momenta of particles,
$p_1$, $p_2$, and $p_3$, expressed in terms of $q_1$,
$q_2$, and $q_3$}
\end{table}

\subsection{Regularization}
\label{sec:regularization}

The first cutoff we introduce is the small $p^+$ cutoff
in the Fourier expansions of fields,
\beq
\psi(x)
\es
\sum_{\sigma} \int\frac{dp^+ d^2p^\perp}{16\pi^3 p^+} \theta_\epsilon(p^+)
\left[
  u_\sigma(p) e^{-i p x} b_{p \sigma}
+ v_\sigma(p) e^{ i p x} d_{p \sigma}^\dag
\right]
 ,
\\
\phi(x)
\es
\int\frac{dp^+ d^2p^\perp}{16\pi^3 p^+} \theta_\epsilon(p^+)
\left(
  e^{-i p x} a_{p}
+ e^{i p x} a_{p}^\dag
\right)
 ,
\eeq
where $\theta_\epsilon(p^+) = \theta(p^+ - \epsilon^+)$
and $\epsilon^+ > 0$ is the minimal $p^+$ any particle
can have. Therefore,
\beq
\Phi(q)
\es
\frac{
  \theta_\epsilon(q^+) a_{q}
+ \theta_\epsilon(-q^+) a_{-q}^\dag
}{|q^+|}
\ ,
\\
\Psi(q)
\es
\sum_{\sigma}
\frac{
  \theta_\epsilon(q^+) u_{\sigma}(q) b_{q \sigma}
+ \theta_\epsilon(-q^+) v_{\sigma}(-q) d_{-q \sigma}^\dag
}{|q^+|}
\ .
\eeq
The purpose of this cutoff is to remove all zero modes
from the theory and ensure the triviality of the vacuum
state. At the same time the limit $\epsilon^+ \to 0$
is the first limit to be taken.

The canonical Hamiltonian is not well-defined because
local interactions lead to ultraviolet divergences.
Hence, we redefine the interaction Hamiltonian to be,
\beq
H_{\psi^2 \phi}
\es
g
\int[q_1 q_2 q_3]_\epsilon
\,f_{t_r,123}
\,\tdelta_{123}
\,\cN\!\left[ \bar\Psi(-q_1) \Psi(q_2) \Phi(q_3) \right] ,
\label{eq:Hpsi2phiReg}
\eeq
where $t_r$ is the cutoff parameter,
\beq
f_{t_r, 1 \dots n}
\es
e^{-t_r(q_1^- + \dots + q_n^-)^2}
\ ,
\\
{}[q]_\epsilon
\es
[q] \, \theta_\epsilon(|q^+|)
\ ,
\quad
[q_1 q_2]_\epsilon
\rs
[q_1]_\epsilon [q_2]_\epsilon
\text{ etc.}
\eeq
The limit $t_r \to 0$ makes $f_{t_r,123} \to 1$, and
corresponds to the removal of the regularization factor.

The regulator works in the following way. Suppose that
$-q_1^+, -q_2^+, q_3^+ > 0$. The combination of creation
and annihilation operators that arises is $b_{-q_1}^\dag
d_{-q_2}^\dag a_{q_3}$. In other words, a boson with momentum
$p_3 = q_3$ is annihilated, an antifermion with momentum
$p_2 = -q_2$ is created, and a fermion with momentum $p_1
= -q_1$ is created. The regulator becomes $e^{-t_r(-p_1^-
- p_2^- + p_3^-)^2}$ and regulates the difference between
the energy of created particles, $p_1^- + p_2^-$, and the
energy of the annihilated particles, $p_3^-$. All six
nonzero cases can be analyzed in the same way and in each
case $f_{t_r,123}$ regulates the difference between
front-form energy before and after the interaction.

The instantaneous interaction is regularized as if it were
composed of two first-order interaction terms,
\beq
H_{\psi^2 \phi^2}
\es
g^2
\int[q_1 q_2 q_3 q_4]_\epsilon
\,\theta_\epsilon(|q_5^+|)
\,\tdelta_{1234}
\,f_{t_r,135}
\,f_{t_r,24.5}
\ \cN\!\left[
\bar\Psi(-q_1) \Phi(q_3)
\frac{\gamma^+}{2 q_5^+}
\Phi(q_4) \Psi(q_2)
\right] ,
\label{eq:Hpsi2phi2reg}
\eeq
where
\beq
f_{t_r,1 \dots k.(k+1) \dots n}
\es
e^{-t_r\left[(q_1^- + \dots + q_k^-)
- (q_{k+1}^- + \dots + q_n^-)\right]^2}
\ ,
\label{eq:defFFdot}
\eeq
and $q_5 = q_2 + q_4 = - q_1 - q_3$ for plus and transverse
components, while $q_5^- = \left. q_5^- \right|_m =
\frac{m^2 + (q_5^\perp)^2}{q_5^+}$.

Regularization makes the Hamiltonian mathematically
well-defined, but the observables strongly depend on
the cutoff parameters. In the next section we start
the renormalization process by calculating the effective
Hamiltonian of Yukawa theory up to the second order
in the coupling constant.

\subsection{Effective Hamiltonian}
\label{sec:effHam}

The first-order Eq.~(\ref{eq:RGPEP1st}) has a simple solution,
\beq
h_{t,1}
\es
h_{0,1}
\rs
H_{\psi^2 \phi}
\ .
\label{eq:ht1}
\eeq
Therefore,
\beq
\cH_{t,1}
\es
g
\int[q_1 q_2 q_3]_\epsilon
\, f_{t_r,123}
\, \tdelta_{123}
\, f_{t,123}
\, \cN\!\left[ \bar\Psi(-q_1) \Psi(q_2) \Phi(q_3) \right] .
\label{eq:YukawaHt1}
\eeq
The interaction term acquires an effective form factor
$f_{t,123}$. Due to the choice of the regulator, the form
factor and the regulator can be combined,
\beq
f_{t,123} f_{t_r,123} \es f_{t + t_r,123} \ .
\eeq
Therefore, the RGPEP equation in the first order of perturbative
calculations lowered the energy cutoff from $1/\sqrt{t_r}$
at $t = 0$ to $1/\sqrt{t + t_r}$ at $t > 0$. Orders higher than
the first lead to more complicated refinements to effective
interactions. Since the form factor has the same form as the
regulator, the latter is no longer needed to prevent divergent
integrals and the limit $t_r \to 0$ can be easily performed.
One can consider regularization in this case as being a form
of preconditioning of the renormalization procedure.

At second order we first note that
\beq
i \pd_f^- \cH_{t,1}
\es
g
\int[q_1 q_2 q_3]_\epsilon
\left( q_1^- + q_2^- + q_3^- \right)
\tdelta_{123}
\, f_{t + t_r,123}
\, \cN\!\left[ \bar\Psi(-q_1) \Psi(q_2) \Phi(q_3) \right] .
\eeq
The left hand side of Eq.~(\ref{eq:RGPEP2nd}) contains
$(-i\pd_f^- \cH_{t,1}) \cH_{t,1} + \cH_{t,1} i\pd_f^- \cH_{t,1}$.
For both terms we assign 1, 2, and 3 as dummy labels for
the left copy of $\cH_{t,1}$, while 4, 5, and 6 are the
dummy labels of the right copy of $\cH_{t,1}$. Therefore,
the second-order Eq.~(\ref{eq:RGPEP2nd}) takes a concise form,
\beq
\frac{d}{dt} h_{t,2}
\es
g^2
\int[q_1 q_2 q_3 q_4 q_5 q_6]_\epsilon
\, f_{t_r,123} f_{t_r,456}
\, \tdelta_{123}
\, \tdelta_{456}
\, A_{t,123.456}
% \\
% \times
\,
\cN\!\left[ \bar\Psi(-q_1) \Psi(q_2) \Phi(q_3) \right]
\cN\!\left[ \bar\Psi(-q_4) \Psi(q_5) \Phi(q_6) \right] ,
\label{eq:dhdt2ndYukawa}
\eeq
where
\beq
A_{t,123.456}
\es
\left( - q_1^- - q_2^- - q_3^- + q_4^- + q_5^- + q_6^- \right)
\frac{ f_{t,123} f_{t,456} }{ f_{t,123456} }
\ .
\label{eq:At}
\eeq
$1/f_{t,123456} = f_{-t,123456}$ comes from $e^{t(i\pd_f^-)^2}$.
The form of Eq.~(\ref{eq:dhdt2ndYukawa}) is in fact very concise.
Once Eq.~(\ref{eq:dhdt2ndYukawa}) is normal ordered, the nonzero
terms produce twenty two distinct types of terms. Dealing with
them separately would be cumbersome.

Integrating Eq.~(\ref{eq:dhdt2ndYukawa}) over $t$ gives,
\begin{multline}
h_{t,2}
=
h_{0,2}
+
g^2
\int[q_1 q_2 q_3 q_4 q_5 q_6]_\epsilon
\, f_{t_r,123} f_{t_r,456}
\, \tdelta_{123}
\, \tdelta_{456}
\, B_{t,123.456}
\\
\times
\cN\!\left[ \bar\Psi(-q_1) \Psi(q_2) \Phi(q_3) \right]
\cN\!\left[ \bar\Psi(-q_4) \Psi(q_5) \Phi(q_6) \right] ,
\label{eq:ht2}
\end{multline}
where
\beq
B_{t,123.456}
\es
\int_0^t d\tau A_{\tau,123.456}
\\
\es
\frac{1}{2}
\left(
  \frac{1}{q_1^- + q_2^- + q_3^-}
- \frac{1}{q_4^- + q_5^- + q_6^-}
\right)
\left(
  \frac{ f_{t,123} f_{t,456} }{ f_{t,123456} }
- 1
\right) .
\label{eq:Bt}
\eeq
The initial condition, $h_{0,2}$ consists of the canonical
terms of order $g^2$ -- in this case the instantaneous fermion
interaction -- and, yet to be determined, counterterms $X$:
\beq
h_{0,2} \es H_{\psi^2 \phi^2} + X \ .
\eeq

To determine the counterterms we need to evaluate the matrix
elements of the effective Hamiltonian. This is most easily done
when the interaction terms are rewritten in a normal-ordered
form. In the next section we describe how to do this using
Wick's theorem and introduce the Wick's diagrams -- a tool
for efficient calculations.

\subsection{Wick's diagrams and Wick's theorem}
\label{sec:wick}

In this section we first introduce Wick's diagrams and
then state the Wick's theorem that is relevant for
our calculations. However, before we introduce the rules
for drawing and interpreting the diagrams, it is important
to understand that our drawings are context-dependent.
The same diagram can correspond to a term in the canonical
Hamiltonian or to a term in the regularized or renormalized
Hamiltonian. Individual diagrams can be composed to
represent higher-order terms contributing to the effective
Hamiltonian or to its derivative. The purpose of the
diagrams is to streamline calculations of the effective
Hamiltonians. A complicated expression for a term in the
Hamiltonian can be represented unambiguously with a simple
diagram. Diagrammatic representation of mathematical formulae
is typically easier to handle for a human calculator. Therefore,
one may expect advantages in organizing and following calculations,
as well as solving the combinatorial problem of generating
all relevant terms (diagrams). Finally, the foundation for
the approach of calculating the effective Hamiltonian using
diagrams is Wick's theorem. Should the proposed rules prove
insufficient to unambiguously generate a mathematical
expression, one should write out the formulas and follow
the theorem.

\begin{figure}
\includegraphics{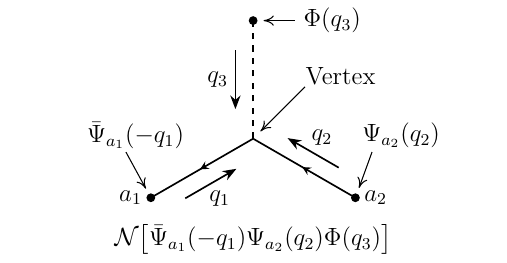}
\caption{\label{fig:YukawaWick1}Wick's diagram of the
first-order Yukawa Hamiltonian term. The vertex function
is $g \delta_{a_1,a_2}$, $g f_{t_r,123} \delta_{a_1,a_2}$,
$g f_{t + t_r,123} \delta_{a_1,a_2}$, depending whether
the diagram represents a term in the canonical,
Eq.~(\ref{eq:Hpsi2phi}), the regularized, Eq.~(\ref{eq:Hpsi2phiReg}),
or the effective Hamiltonian, Eq.~(\ref{eq:YukawaHt1}),
respectively. The orientation of the legs is arbitrary.}
\end{figure}

The simplest Wick's diagram in Yukawa theory corresponds to
the first-order term in the Hamiltonian and is depicted
in Fig.~\ref{fig:YukawaWick1}. We start by drawing the
legs of the diagram -- dashed line for a boson, two directed
solid lines for fermions. With each leg we associate
momentum variable, $q_1$, $q_2$, and $q_3$. The momentum
flow is directed as indicated by the additional arrows.
We place dots, which correspond to field operators, at
the loose ends of the legs. The dots on fermion lines are
labeled with spinor indices, $a_1$ and $a_2$. The dot at
the end of the dashed line corresponds to $\Phi(q_3)$.
The dot at the end of the solid line which is directed away
from the dot corresponds to $\Psi_{a_2}(q_2)$, while the dot
at the end of the solid line which is directed toward the dot
corresponds to $\bar\Psi_{a_1}(-q_1)$. If the direction of
any momenta were inverted, then the momentum in the corresponding
operator would also need to be inverted (replaced with
its negative). Once all operators are determined, they
are written in some canonical order. For example, first
fermion fields following the solid line in the direction
opposite to the direction on the line, then boson fields,
finally they are all normal ordered. Therefore, we write
$\cN\left[\bar\Psi_{a_1}(-q_1) \Psi_{a_2}(q_2) \Phi(q_3)\right]$.
The order in which we write the operators matters only
for the fermionic operators, because $\cN\left[
\bar\Psi_{a_1}(-q_1) \Psi_{a_2}(q_2)\right] = -\cN\left[
\Psi_{a_2}(q_2) \bar\Psi_{a_1}(-q_1)\right]$.

The intersection of the three lines in Fig.~\ref{fig:YukawaWick1},
called a vertex, corresponds to a function called the
vertex function. The vertex function depends on the context.
If the diagram represents a term in the canonical Hamiltonian,
then the vertex function is equal to the coupling constant
times Kronecker delta in spinor indices, $g \delta_{a_1,a_2}$
and might be called a vertex constant in this case. If the
diagram represents a term in the regularized Hamiltonian,
then the vertex function equals $g f_{t_r,123} \delta_{a_1,a_2}$
and is a function of momenta associated with the legs
of the diagram. If the diagram represents a term in the
effective Hamiltonian, then the vertex function equals
$g f_{t + t_r,123} \delta_{a_1,a_2}$. Each vertex also
comes with a momentum conservation Dirac delta, which is
$\tdelta_{123}$ in this case.

Finally, all labels need to be integrated or summed over.
The spinor indices $a_1$ and $a_2$ are summed over 1, 2, 3,
and 4, each. Throughout this article we assume Einstein
summation convention, hence, in any formula that shows
the spinor indices, the summation over spinor indices
is implicit. Momentum variables need to be integrated
with measure $[q_1 q_2 q_3]_\epsilon$. When all factors
are combined, we arrive at Eq.~(\ref{eq:YukawaHt1}), the
Hamiltonian term corresponding to Fig.~\ref{fig:YukawaWick1}.

\begin{figure}
\includegraphics{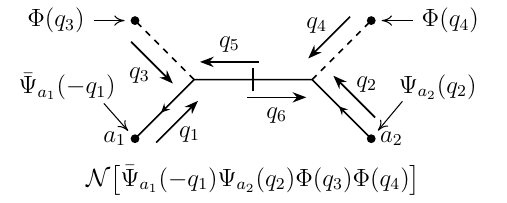}
\caption{\label{fig:YukawaWick2inst}Wick's diagram of
the second-order instantaneous fermion interaction. It
is interpreted as two first-order vertices connected
to each other. The vertex function for the canonical
Hamiltonian is $g^2 \gamma^+_{a_1,a_2} / (2 q_5^+)$,
see Eq.~(\ref{eq:Hpsi2phi2}). For the regularized
Hamiltonian it is $g^2 f_{t_r,135} f_{t_r,246}
\gamma^+_{a_1,a_2} / (2 q_5^+)$, see
Eq.~(\ref{eq:Hpsi2phi2reg}), and for the effective
Hamiltonian it is $g^2 f_{t_r,135} f_{t_r,246} f_{t,1234}
\gamma^+_{a_1,a_2} / (2 q_5^+)$, see $\cH_{t,2}^{(0)}$
defined immediately before Eq.~(\ref{eq:cHt0bc}).
Legs can be oriented in arbitrary direction.}
\end{figure}

In the second-order calculation, the reduced Hamiltonian
comprises two terms, see Eq.~(\ref{eq:ht2}). $h_{0,2}$
contains $H_{\psi^2 \phi^2}$, which is depicted in
Fig.~\ref{fig:YukawaWick2inst}. We interpret this second-order
diagram as two first-order diagrams connected with each other.
Applying the previous rules, we get the following. There are
four dots, hence, there are four field operators, in the order
we described before: $\bar\Psi_{a_1}(-q_1)$, $\Psi_{a_2}(q_2)$,
$\Phi(q_3)$, and $\Phi(q_4)$. Therefore, the first part of
the expression is $\cN\left[\bar\Psi_{a_1}(-q_1) \Psi_{a_2}(q_2)
\Phi(q_3) \Phi(q_4) \right]$. Two vertices imply two Dirac deltas
for momentum conservation, $\tdelta_{135}$ and $\tdelta_{246}$.
The integration measure is $[q_1 q_2 q_3 q_4 q_5 q_6]_\epsilon$.

Going beyond the rules established so far we add a Dirac delta
$\tdelta_{56}$ because the intermediate line that connects the
two vertices can be thought of as two lines connected together.
The second-order vertex as a whole is assigned a single vertex
function. To reproduce Eq.~(\ref{eq:Hpsi2phi2reg}) the vertex
function needs to be $g^2 f_{t_r,135} f_{t_r,246} \gamma^+_{a_1,a_2}
/ (2 q_5^+)$. To reproduce Eq.~(\ref{eq:Hpsi2phi2}) the vertex
function needs to be $g^2 \gamma^+_{a_1,a_2} / (2 q_5^+)$.
If the diagram is to represent the contribution to the effective
Hamiltonian, then the vertex function is $g^2 f_{t_r,135} f_{t_r,246}
f_{t,1234} \gamma^+_{a_1,a_2} / (2 q_5^+)$.

The three Dirac deltas
allow us to perform some integrations easily. Firstly, one can
integrate over $q_6$, which will remove $\tdelta_{56}$, substitute
every $q_6$ with $-q_5$, and leave $\theta_\epsilon(|q_5^+|)$
behind from the measure. One can then integrate over $q_5$,
which will remove $\tdelta_{24.5}$. For convenience, one can
keep using $q_5$ as a place holder for $q_2 + q_4$. Integration
leaves another $\theta_\epsilon(|q_5^+|)$ behind from the measure,
and $\theta_\epsilon(|q_5^+|)^2 = \theta_\epsilon(|q_5^+|)$.
One could take those simplifications into account by defining
special rules for the vertex, assuming that the internal line
represents a single, extended vertex (imagine a single point
stretched to a segment of a line). For example, we can assign
the overall Dirac delta $\tdelta_{1234}$, integration measure
$[q_1 q_2 q_3 q_4]_\epsilon \theta_\epsilon(|q_5^+|)$, and
the same appropriate vertex function as stated before, assuming
$q_5 = q_2 + q_4 = -q_6$.

The second term in Eq.~(\ref{eq:ht2}) introduces new, effective
interactions to the Hamiltonian. To simplify the expression, we
use Wick's theorem. For that purpose we need to introduce
contractions of the fields. A contraction is defined for two
objects that are proportional to creation or annihilation
operators, such as $\Phi$ and $\Psi$. Given two such operators,
$A$ and $B$, the contraction (due to Houriet and
Kind~\cite{hourietkind}), $\wick{\c{A}\c{B}} = A B - \cN(A B)$.
Hence, field contractions are,
\beq
\wick{
\c{\Psi}_{a_1}(q_1)
\,
\c{\bar\Psi}_{a_2}(-q_2)
}
\es
+
\frac{ \theta_\epsilon(q_1^+) }{ q_1^+ }
\, \tdelta_{12}
\ ( \slashed{q}_1 + m )_{a_1, a_2}
\ ,
\label{eq:contractionPsiPsi1}
\\
\wick{
\c{\bar\Psi}_{a_2}(-q_2)
\,
\c{\Psi}_{a_1}(q_1)
}
\es
-
\frac{ \theta_\epsilon(q_2^+) }{ q_2^+ }
\, \tdelta_{12}
\ (\slashed q_1 + m)_{a_1, a_2}
\ ,
\label{eq:contractionPsiPsi2}
\\
\wick{
\c{\Psi}_{a_1}(q_1)
\,
\c{\Psi}_{a_2}(q_2)
}
\es
\wick{
\c{\bar\Psi}_{a_1}(-q_1)
\,
\c{\bar\Psi}_{a_2}(-q_2)
}
\rs
0
\ ,
\phantom{\frac{|}{|}}
\\
\wick{
\c{\Phi}(q_1)
\,
\c{\Phi}(q_2)
}
\es
\frac{ \theta_\epsilon(q_1^+) }{ |q_1^+| }
\,\tdelta_{12}
\ .
\label{eq:contractionPhiPhi}
\eeq

Wick's theorem states that a product of operators such as
$A_1 \dots A_n$ is equal to the normal-ordered sum of all
possible terms with multiple contractions including zero
contractions in a term, one contraction in a term, two
contractions, three, etc. The most contractions possible
in the example is $\lfloor n/2 \rfloor$. Therefore, the sum
contains terms such as $\cN(A_1 \dots A_n)$,
$\wick{\cN(\c{A_1} \c{A_2} \dots A_n)}$,
$\wick{\cN(A_1 \c{A_2} A_3 \dots \c{A_n})}$,
$\wick{\cN(\c{A_1} \c{A_2} \c{A_3} \c{A_4} \dots A_n)}$,
$\wick{\cN(\c1{A_1} \c2{A_2} \c1{A_3} \c2{A_4} \dots A_n)}$,
etc. To compute contractions between operators that are not
adjacent one needs to transpose operators until the two
operators are adjacent. Whenever both transposed operators
are fermionic, one needs to add an overall factor of $-1$
to that term. Importantly, one should never transpose two
operators that are contracted with each other because
$\wick{\c{A} \c{B}} \neq \wick{\c{B} \c{A}}$.

For our purposes, we need to use the second theorem by
Wick~\cite{Wick:1950ee}. Expressions that are of relevance for
us are of the form $\cN(A_1 \dots A_k) \cN(A_{k+1} \dots A_n)$,
cf. Eq.~(\ref{eq:ht2}).
In this case, we have to write down the same kind of sum that
expresses the first Wick's theorem for $A_1 \dots A_n$ except
that every contraction line has to start on one of $A_1$, \dots,
$A_k$ operators, and end on one of $A_{k+1}$, \dots, $A_n$
operators. Contractions between operators in the set $A_1$,
\dots, $A_k$ are forbidden because they are enclosed within
normal ordering symbol, hence one can transpose them ``for free,''
i.e., without producing a contraction. For example, $a a^\dag
= a^\dag a + \wick[offset=9pt]{\c{a} \c{a}^\dag}$, but
$\cN(a a^\dag) = \cN(a^\dag a)$.

\begin{figure}
\includegraphics{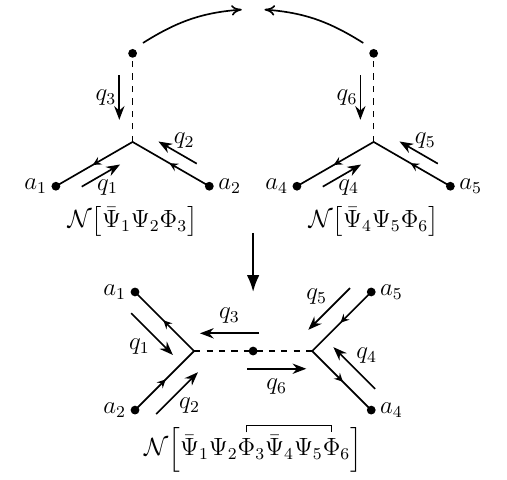}
\caption{\label{fig:contraction}Wick contraction is
represented by merging dots corresponding to the
contracted operators. External legs can be oriented
arbitrarily, but the left-right order of vertices
cannot be changed, because changing the order of the
contracted operators changes the resulting contraction.
Notation is simplified, $\bar\Psi_1 = \bar\Psi(-q_1)$,
$\Psi_2 = \Psi(q_2)$, $\Phi_3 = \Phi(q_3)$, etc.}
\end{figure}

\begin{figure*}
\includegraphics{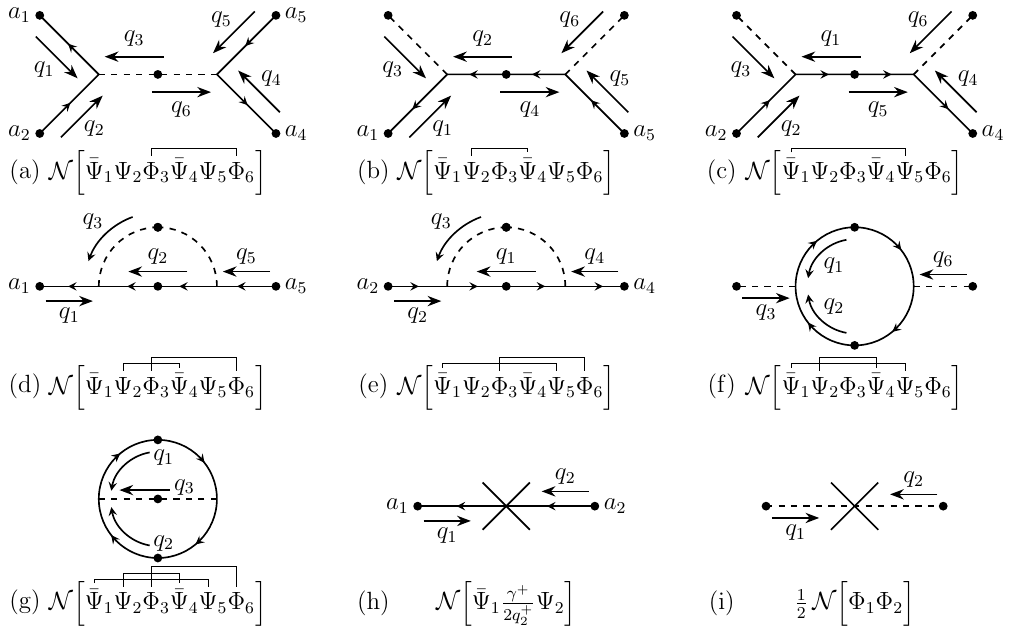}
\caption{\label{fig:YukawaWick2}Second-order Wick's
diagrams. Diagrams (a) through (g) correspond to Wick
contractions, Eq.~(\ref{eq:yukawa2NN}). Diagrams (h)
and (i) represent mass counterterms that need to be
added to the initial, bare Hamiltonian,
Eqs.~(\ref{eq:YukawaMassXF}) and (\ref{eq:YukawaMassXB}).
Notation is simplified, $\bar\Psi_1 = \bar\Psi(-q_1)$,
$\Psi_2 = \Psi(q_2)$, $\Phi_3 = \Phi(q_3)$, etc.}
\end{figure*}

According to Wick's theorem,
\begin{multline}
\cN\!\left[ \bar\Psi(-q_1) \Psi(q_2) \Phi(q_3) \right]
\cN\!\left[ \bar\Psi(-q_4) \Psi(q_5) \Phi(q_6) \right]
\\
=
\cN\!\left[ \bar\Psi(-q_1) \Psi(q_2) \Phi(q_3)
\bar\Psi(-q_4) \Psi(q_5) \Phi(q_6) \right]
+ D_a
+ D_b
+ D_c
+ D_d
+ D_e
+ D_f
+ D_g
\ ,
\label{eq:yukawa2NN}
\end{multline}
where terms $D_a$ through $D_g$ are terms with field
contractions: $D_a$, $D_b$, and $D_c$ have one contraction
each, $D_d$, $D_e$, and $D_f$ have two contractions
each, while $D_g$ has three contractions. For each
case one imagines two first-order diagrams drawn one next
to the other. Contractions are represented by connecting
lines that correspond to the contracted operators, see
Fig.~\ref{fig:contraction}. All seven terms, $D_a$ through
$D_g$ are depicted in Fig.~\ref{fig:YukawaWick2} on panels
(a) through (g), respectively.

The reduced Hamiltonian, $h_{t,2}$ is decomposed into
a sum of terms that directly corresponds to the sum in
Eq.~(\ref{eq:yukawa2NN}),
\beq
h_{t,2}
\es
  h_{0,2}
+ h_{t,2}^{(a)} + h_{t,2}^{(b)}
+ h_{t,2}^{(c)} + h_{t,2}^{(d)}
+ h_{t,2}^{(e)} + h_{t,2}^{(f)}
+ h_{t,2}^{(g)}
\ ,
\label{eq:ht2decomp}
\\
h_{t,2}^{(J)}
\es
g^2
\int[q_1 q_2 q_3 q_4 q_5 q_6]_\epsilon
\, f_{t_r,123} f_{t_r,456}
\, \tdelta_{123}
\, \tdelta_{456}
\, B_{t,123.456}
\, D_J
\ ,
\eeq
where $J \in \left\{ a, \dots, g \right\}$.
Equation~(\ref{eq:ht2decomp}) does not contain a term
that corresponds to the first term on the right hand
side in Eq.~(\ref{eq:yukawa2NN}) because that term gives zero
contribution after integration in Eq.~(\ref{eq:ht2})
-- it is symmetric with respect to the change of labels
$1,2,3 \leftrightarrow 4,5,6$ (inside normal-ordering
parentheses one can freely transpose field operators
as long as one keeps track of the overall sign), while
$B_{t,123.456}$ is antisymmetric.\footnote{ In Eq.~(\ref{eq:ht2}),
similar argument does not work because $\cN\!\left[
\bar\Psi(-q_1) \Psi(q_2) \Phi(q_3) \right]$ and
$\cN\!\left[ \bar\Psi(-q_4) \Psi(q_5) \Phi(q_6)
\right]$ do not commute.} This is a general result,
due to a commutator in RGPEP Eq.~(\ref{eq:RGPEP})
only terms with nonzero number of contractions
contribute to the effective Hamiltonians.

The $D_g$ term, where
\beq
D_g
\es
\cN\!\left[ \wick{
\c1{\bar\Psi}(-q_1) \c2{\Psi}(q_2) \c3{\Phi}(q_3)
\c2{\bar\Psi}(-q_4) \c1{\Psi}(q_5) \c3{\Phi}(q_6)
} \right] ,
\eeq
leads to $h_{t,2}^{(g)} = 0$ because momentum
conservation requires $q_1^+ + q_2^+ + q_3^+ = 0$,
while $q_1^+, q_2^+, q_3^+ \ge \epsilon^+ > 0$ for
the cutoff Hamiltonian.

Using Wick's diagrams, we discuss expressions for
terms $D_a$, $D_b$, and $D_c$ in Sec.~\ref{sec:H2tree},
and expressions for terms $D_d$, $D_e$, and $D_f$
in Sec.~\ref{sec:H2self}.

\subsection{\label{sec:H2tree}Tree diagrams}

In this section we discuss terms of the effective
second-order Hamiltonian that can be represented
with tree diagrams, i.e., single-contraction terms
$D_a$, $D_b$, and $D_c$, and their corresponding
reduced Hamiltonians, $h_{t,2}^{(a)}$, $h_{t,2}^{(b)}$,
and $h_{t,2}^{(c)}$.

The $D_a$ term is,
\beq
D_a
\es
\cN\!\left[ \wick{
\bar\Psi(-q_1) \Psi(q_2) \c{\Phi}(q_3)
\bar\Psi(-q_4) \Psi(q_5) \c{\Phi}(q_6)
} \right] .
\eeq
This term is depicted in Fig.~\ref{fig:YukawaWick2}(a). Four
external dots represent field operators $\bar\Psi(-q_1)$,
$\Psi(q_2)$, $\bar\Psi(-q_4)$, and $\Psi(q_5)$. The dot
on the internal line is obtained by merging two dots
corresponding to two scalar field operators, see
Fig.~\ref{fig:contraction}. Therefore, the internal dot
represents contraction $\wick{\c{\Phi}(q_3) \c{\Phi}(q_6)}$.
In order to reproduce $h_{t,2}^{(a)}$ from the diagram,
one first writes all operators of the left vertex,
$\bar\Psi_{a_1}(-q_1) \Psi_{a_2}(q_2) \Phi(q_3)$, then
the operators of the right vertex, $\bar\Psi_{a_4}(-q_4)
\Psi_{a_5}(q_5) \Phi(q_6)$ are placed to the right of the
left vertex operators, and then one adds necessary contractions.
With each vertex comes the momentum conservation Dirac delta
and the vertex function of the regularized Hamiltonian. Each
leg introduces a momentum integration. The integration measure
is $[q_1 q_2 q_3 q_4 q_5 q_6]_\epsilon$. If the diagram
corresponds to a term in the reduced Hamiltonian then an overall
function, $B_{t,123.456}$ is assigned to the diagram. If the
diagram corresponds to a term in the effective Hamiltonian,
then the overall factor that needs to be assigned is
$f_{t,123456} B_{t,123.456}$.

Using Eq.~(\ref{eq:contractionPhiPhi}) and integrating
over $q_6$, the reduced Hamiltonian term simplifies to
\beq
h_{t,2}^{(a)}
\es
g^2
\int[q_1 q_2 q_3 q_4 q_5]_\epsilon
\, \tdelta_{1245}
\, \tdelta_{45.3}
\, B_{t,123.45(-3)}
\, \frac{\theta_\epsilon(q_3^+)}{q_3^+}
\cN\!\left[
\bar\Psi(-q_1) \Psi(q_2)
\bar\Psi(-q_4) \Psi(q_5)
\right] ,
\label{eq:ht2a}
\eeq
where $(-3)$ in $B_{t,123.45(-3)}$ means that momentum $-q_3$
is used as an argument. Explicitly:
\beq
B_{t,123.45(-3)}
\es
\frac{1}{2}
\left(
  \frac{1}{q_1^- + q_2^- + q_3^-}
- \frac{1}{q_4^- + q_5^- - q_3^-}
\right)
\left(
  \frac{ f_{t,123} f_{t,45.3} }{ f_{t,1245} }
- 1
\right) ,
\eeq
where we used, in accordance with Eq.~(\ref{eq:defFFdot}),
$f_{t,45.3} = f_{t,45(-3)}$, and $f_{t,12345(-3)}
= f_{t,1245}$.

The $D_b$ and $D_c$ terms, depicted in
Figs.~\ref{fig:YukawaWick2}(b) and
\ref{fig:YukawaWick2}(c), respectively,
are
\beq
D_b
\es
\cN\!\left[ \wick{
\bar\Psi(-q_1) \c{\Psi}(q_2) \Phi(q_3)
\c{\bar\Psi}(-q_4) \Psi(q_5) \Phi(q_6)
} \right] ,
\\
D_c
\es
\cN\!\left[ \wick{
\c{\bar\Psi}(-q_1) \Psi(q_2) \Phi(q_3)
\bar\Psi(-q_4) \c{\Psi}(q_5) \Phi(q_6)
} \right] .
\eeq
We present a step by step evaluation of
Fig.~\ref{fig:YukawaWick2}(c), which should serve
a good illustration of all elements of the calculations
one needs to perform in the second order. In the first
step we write down the operators and contractions based
on Fig.~\ref{fig:YukawaWick2}(c):
\beq
\text{Step 1:}&\quad&
\cN\!\left[ \wick{
\c{\bar\Psi}_{a_1}(-q_1) \Psi_{a_2}(q_2) \Phi(q_3)
\bar\Psi_{a_4}(-q_4) \c{\Psi}_{a_5}(q_5) \Phi(q_6)
} \right]
\eeq
In the second step we add vertex functions
of the regularized Hamiltonian:
\beq
\text{Step 2:}&\quad&
g^2
f_{t_r,123}
f_{t_r,456}
\delta_{a_1,a_2}
\delta_{a_4,a_5}
% \nt
\cN\!\left[ \wick{
\c{\bar\Psi}_{a_1}(-q_1) \Psi_{a_2}(q_2) \Phi(q_3)
\bar\Psi_{a_4}(-q_4) \c{\Psi}_{a_5}(q_5) \Phi(q_6)
} \right]
\eeq
In the third step we evaluate the contraction using
Eq.~(\ref{eq:contractionPsiPsi2}):
\beq
\text{Step 3:}&\quad&
-g^2
f_{t_r,123}
f_{t_r,456}
\delta_{a_1,a_2}
\delta_{a_4,a_5}
\frac{ \theta_\epsilon(q_1^+) }{ q_1^+ }
\, \tdelta_{51}
\ (\slashed q_5 + m)_{a_5, a_1}
% \nt
\cN\!\left[
\Psi_{a_2}(q_2) \Phi(q_3)
\bar\Psi_{a_4}(-q_4) \Phi(q_6)
\right]
\eeq
In the fourth step we continue simplifying. We can reorder
the operators producing a minus sign. Additionally, we put
$\slashed{q}_5 + m$ between the fermionic operators and
simplify the summation over $a_5$ and $a_1$:
\beq
\text{Step 4:}&\quad&
+g^2
f_{t_r,123}
f_{t_r,456}
\frac{ \theta_\epsilon(q_1^+) }{ q_1^+ }
\, \tdelta_{51}
% \nt
\cN\!\left[
\bar\Psi_{a_4}(-q_4)
(\slashed q_5 + m)_{a_4, a_2}
\Psi_{a_2}(q_2) 
\Phi(q_3) \Phi(q_6)
\right]
\eeq
In this form one can drop the spinor indices $a_4$,
and $a_2$ because $\bar\Psi$ is a row array,
$\slashed{q}_5 + m$ is a matrix, and $\Psi$ is
a column vector.

The fifth step is to add the $B_{t,123.456}$ factor
defined in Eq.~(\ref{eq:Bt}), vertex momentum
conservation deltas, and integrations:
\begin{multline}
\text{Step 5:}\quad
h_{t,2}^{(c)}
=
g^2
\int[q_1 q_2 q_3 q_4 q_5 q_6]_\epsilon
B_{t,123.456}
\tdelta_{123}
\tdelta_{456}
f_{t_r,123}
f_{t_r,456}
\frac{ \theta_\epsilon(q_1^+) }{ q_1^+ }
\, \tdelta_{51}
\\ \times
\cN\!\left[
\bar\Psi(-q_4) (\slashed q_5 + m) \Psi(q_2) 
\Phi(q_3) \Phi(q_6)
\right] .
\label{eq:ht2c}
\end{multline}
In a similar manner, Fig.~\ref{fig:YukawaWick2}(b)
evaluates to
\beq
h_{t,2}^{(b)}
\es
g^2
\int[q_1 q_2 q_3 q_4 q_5 q_6]_\epsilon
\, B_{t,123.456}
\, \tdelta_{123}
\, \tdelta_{456}
f_{t_r,123}
f_{t_r,456}
\frac{ \theta_\epsilon(q_2^+) }{ q_2^+ }
\, \tdelta_{24}
% \\
% \times
\,
\cN\!\left[ \wick{
\bar\Psi(-q_1) (\slashed q_2 + m) \Psi(q_5)
\Phi(q_3) \Phi(q_6)
} \right] .
\label{eq:ht2b}
\eeq
Figures \ref{fig:YukawaWick2}(b) and \ref{fig:YukawaWick2}(c)
are similar and in fact they complement each other. To show
this, one takes $h_{t,2}^{(c)}$, renames labels
$1 \leftrightarrow 4$, $2 \leftrightarrow 5$,
$3 \leftrightarrow 6$, transpose the two $\Phi$ fields, and
use the fact that $B_{t,456.123} = - B_{t,123.456}$. Next,
in both $h_{t,2}^{(b)}$ and $h_{t,2}^{(c)}$, we perform
integration over $q_4$, which leads to the two becoming
almost identical. The only difference then is that $h_{t,2}^{(b)}$
contains $\theta_\epsilon(q_2^+)$, while $h_{t,2}^{(c)}$ contains
$\theta_\epsilon(-q_2^+)$. Since $\theta_\epsilon(q_2^+)
+ \theta_\epsilon(-q_2^+) = \theta_\epsilon(|q_2^+|)$,
the two expressions can be combined into,
\begin{multline}
h_{t,2}^{(b)} + h_{t,2}^{(c)}
=
g^2
\int[q_1 q_2 q_3 q_5 q_6]_\epsilon
\, B_{t,123.(-2)56}
\, \tdelta_{1356}
\, \tdelta_{56.2}
f_{t_r,123}
f_{t_r,56.2}
\frac{ \theta_\epsilon(|q_2^+|) }{ q_2^+ }
\\
\times
\cN\!\left[ \wick{
\bar\Psi(-q_1) \Phi(q_3)
(\slashed q_2 + m)
\Phi(q_6) \Psi(q_5)
} \right] ,
\label{eq:ht2bc}
\end{multline}
where for aesthetic reasons the two $\Phi$ fields are reordered
again.

The simplification of the sum of Figs.~\ref{fig:YukawaWick2}(b)
and \ref{fig:YukawaWick2}(c) is a manifestation of a symmetry.
Figure \ref{fig:YukawaWick2}(c) can be obtained from
Fig.~\ref{fig:YukawaWick2}(b) by transposing the left and
right vertex or alternatively by changing the direction of
the fermionic line. Therefore, both diagrams are symmetric
with respect to the two combined transformations. This suggests
that there might be a way to define a different kind of
diagram for which Fig.~\ref{fig:YukawaWick2}(b) and
Fig.~\ref{fig:YukawaWick2}(c) are combined into one and
the ordering of the vertices does not matter. Such extensions
may be very useful in calculations in orders higher than
the second, but in the current context they seem to be
form over substance and are not pursued in this work.

\subsection{\label{sec:H2self}Self-energy interactions}

Terms, $D_d$, $D_e$, and $D_f$ in Eq.~(\ref{eq:yukawa2NN}),
all have two contractions and are represented with diagrams
with a loop, see Fig.~\ref{fig:YukawaWick2}. They contribute
to self energies of the fermions and bosons. The fermionic
contributions, depicted in Figs.~\ref{fig:YukawaWick2}(d)
and \ref{fig:YukawaWick2}(e), are
\beq
D_d
\es
\cN\!\left[ \wick{
\bar\Psi(-q_1) \c1{\Psi}(q_2) \c2{\Phi}(q_3)
\c1{\bar\Psi}(-q_4) \Psi(q_5) \c2{\Phi}(q_6)
} \right] ,
\\
D_e
\es
\cN\!\left[ \wick{
\c1{\bar\Psi}(-q_1) \Psi(q_2) \c2{\Phi}(q_3)
\bar\Psi(-q_4) \c1{\Psi}(q_5) \c2{\Phi}(q_6)
} \right] .
\eeq
Figure \ref{fig:YukawaWick2}(d) is evaluated in full analogy
with Fig.~\ref{fig:YukawaWick2}(b). The result is,
\beq
h_{t,2}^{(d)}
\es
g^2
\int[q_1 q_2 q_3]_\epsilon
\, \tdelta_{123}
\, \frac{\theta_\epsilon(q_3^+)}{q_3^+}
\, \frac{\theta_\epsilon(q_2^+)}{q_2^+}
\, f_{t_r,123}^2
\, B_{t,123.(-1)(-2)(-3)}
\,\cN\!\left[
\bar\Psi(-q_1)
(\slashed q_2 + m)
\Psi(-q_1)
\right] ,
% \nn
\label{eq:ht2d}
\eeq
where we used $q_4 = -q_2$, $q_5 = -q_1$, and $q_6 = -q_3$,
which follow from conservation of momentum. Additionally,
symmetries such as $f_{t,(-1)(-2)(-3)} = f_{t,123}$, and
$B_{t,213.456} = B_{t,123.456}$ were also used. Even though
there are two contractions, hence, two internal lines, there
is only one overall factor $B_{t,123.456}$ that is associated
with the diagram. Note also that $q_2^+ > 0$ and $q_3^+ > 0$
imply $q_1^+ < 0$. Therefore, $\bar\Psi(-q_1) \Psi(-q_1)$
is proportional to $b_1^\dag b_1$. Hence,
Fig.~\ref{fig:YukawaWick2}(d) depicts the self energy term
for the fermion.

Figure \ref{fig:YukawaWick2}(e) is evaluated in analogy
to Fig.~\ref{fig:YukawaWick2}(c). Momentum conservation
implies the same relations between momenta as those for
Fig.~\ref{fig:YukawaWick2}(d). The result:
\beq
h_{t,2}^{(e)}
\es
g^2
\int[q_1 q_2 q_3]_\epsilon
\, \tdelta_{123}
\, \frac{\theta_\epsilon(q_3^+)}{q_3^+}
\, \frac{\theta_\epsilon(q_1^+)}{q_1^+}
\, f_{t_r,123}^2
\, B_{t,123.(-1)(-2)(-3)}
\,\cN\!\left[
\bar\Psi(q_2)
(-\slashed q_1 + m)
\Psi(q_2)
\right] .
% \nn
\label{eq:ht2e}
\eeq
$q_1^+ > 0$ and $q_3^+ > 0$ imply $q_2^+ < 0$, which
means that $\bar\Psi(q_2) \Psi(q_2)$ is proportional
to $d_2 d_2^\dag$. Therefore, Fig.~\ref{fig:YukawaWick2}(e)
depicts the self energy term for the antifermion.

The two diagrams can be combined into one expression.
If in Fig.~\ref{fig:YukawaWick2}(d) we make a substitution
$q_1 \to -q_1$, while in Fig.~\ref{fig:YukawaWick2}(e) we
substitute $q_1 \to -q_2$, $q_2 \to q_1$, and $q_3 \to -q_3$,
then
\begin{multline}
h_{t,2}^{(d)} + h_{t,2}^{(e)}
=
g^2
\int[q_1 q_2 q_3]_\epsilon
\, \tdelta_{23.1}
\, \frac{ B_{t,(-1)23.1(-2)(-3)} }{ q_2^+  q_3^+ }
\, f_{t_r,23.1}^2
\,\cN\!\left[
\bar\Psi(q_1)
(\slashed q_2 + m)
\Psi(q_1)
\right]
\\
\times
\left[
  \theta_\epsilon(q_2^+) \theta_\epsilon(q_3^+)
- \theta_\epsilon(-q_2^+) \theta_\epsilon(-q_3^+)
\right] ,
\label{eq:YukawaSelfIntF}
\end{multline}
where
\beq
B_{t,(-1)23.1(-2)(-3)}
\es
\frac{f_{t,23.1}^2 - 1}{q_2^- + q_3^- - q_1^-}
\rs
q_1^+
\frac{e^{-2t\frac{\left[(q_2 + q_3)^2 - q_1^2\right]^2}
{(p_1^+)^2}} - 1}
{(q_2 + q_3)^2 - q_1^2}
\ .
\eeq
Using the Gordon identity~\cite{Peskin:1995ev}, we can
simplify and write the self-energy terms in a form
analogous to Eq.~(\ref{eq:yukawaFreeHamiltonian}):
\beq
h_{t,2}^{(d)} + h_{t,2}^{(e)}
\es
\int[q]_\epsilon \frac{\delta m^2}{q^+}
\,\cN\!\left[
    \bar\Psi(q)
    \frac{\gamma^+}{2}
    \Psi(q)
  \right] ,
\label{eq:ht2de}
\eeq
where
\beq
\delta m^2
\es
g^2
\int[q_2 q_3]_\epsilon
\frac{ q_1^+ \tdelta_{23.1} }{ q_2^+ q_3^+ }
\, \frac{f_{t+t_r,23.1}^2 - f_{t_r,23.1}^2}{(q_2 + q_3)^2 - q_1^2}
\, (2 q_1 \cdot q_2 + 2 m^2)
\ ,
\label{eq:YukawaMassShift}
\eeq
with $q_1^2 = m^2$, $(q_2 + q_3)^2 = (q_{2\mu} + q_{3\mu})
(q_2^\mu + q_3^\mu)$, and $q_2 \cdot q_3 = q_{2\mu} q_3^\mu$.
$\delta m^2$ does depend on $q_1^+$. Therefore, the effective
vertex is not strictly speaking a mass, because mass should
not depend on momentum. To understand it one needs to make
several observations. First, $(q_2 + q_3)^2 = \cM_{23}^2$ is
an invariant mass squared of particles 2 and 3, and in the
front form can be expressed using relative momenta only.
Those momenta are $x = q_2^+/(q_2^+ + q_3^+)$, and
$k^\perp = (1-x) q_2^\perp - x q_3^\perp$. In other words,
$\cM_{23}^2 = [m^2 + (k^\perp)^2]/x + [\mu^2 + (k^\perp)^2]/(1-x)$
does not depend on $q_1$. Second,
\beq
2 q_1 \cdot q_2 + 2 m^2
\es
\frac{(k^\perp)^2 + m^2 (1 + x)^2}{x}
\eeq
does not depend on $q_1$. Third,
\beq
f_{t,23.1}
\es
\exp\left[
- \frac{t}{(q_1^+)^2}
\left(\cM_{23}^2 - m^2\right)^2
\right] .
\eeq
Hence, the form factors depend on $q_1$, but only through
ratio $t/(q_1^+)^2$. Fourth,
\beq
\int[q_2 q_3]_\epsilon
\frac{ q_1^+ \tdelta_{23.1} }{ q_2^+ q_3^+ }
\es
\int\frac{dP^+ d^2P^\perp}{16\pi^3 P^+}
\int\frac{dx d^2 k^\perp}{16\pi^3 x (1-x)}
\theta\left( x - \frac{\epsilon^+}{q_1^+} \right)
\theta\left( 1 - x - \frac{\epsilon^+}{q_1^+} \right)
\, q_1^+ 16\pi^3 \delta^3(P^{+,\perp} - q_1^{+,\perp})
\ ,
\eeq
where we made change of variables $q_2^+, q_2^\perp,
q_3^+, q_3^\perp \to x, k^\perp, P^+, P^\perp$,
where $P = q_2 + q_3$. After $P$ is integrated
over, which is trivial to do because of the Dirac
delta function, explicit $q_1$ survives only through
$\epsilon^+/q_1^+$ in the measure and $t/(q_1^+)^2$
and $t_r/(q_1^+)^2$ in the form factors. Since $t_r$
is enough to regulate the integral one can let
$\epsilon^+ \to 0$. Dependence on $t_r/(q_1^+)^2$
has to be eliminated by the counterterm, cf.
Eqs.~(\ref{eq:IFdiv}) and (\ref{eq:mXfermion}). This means
that the final, renormalized vertex will still depend
on $t/(q_1^+)^2$. This is the source of the aforementioned
breaking of the front-form longitudinal boost invariance.

Observables are in principle independent of $t$ by
construction. According to the front-form dimensional
analysis~\cite{Wilson:1994fk}, $t$ has dimension
$(x^\perp)^4 (x^-)^{-2}$ or $\text{mass}^4 (P^+)^2$.
Therefore, every $t$ needs to be divided by longitudinal
momentum squared of some particle because there are no
parameters that carry dimension of $P^+$. Therefore,
changing $t$ is equivalent to rescaling all longitudinal
momenta, which in turn is equivalent to a longitudinal
boost transformation. Therefore, in principle boost
invariance is conserved. In practice, however,
the unitary transformation is not exact when
perturbation theory is used. Therefore, $t$-independence
is only approximate and, hence, boost invariance is
also only approximate.

The scalar self energy is represented
in Fig.~\ref{fig:YukawaWick2}(f), and
\beq
D_f
\es
\cN\!\left[ \wick{
\c1{\bar\Psi}(-q_1) \c2{\Psi}(q_2) \Phi(q_3)
\c2{\bar\Psi}(-q_4) \c1{\Psi}(q_5) \Phi(q_6)
} \right] .
\eeq
Evaluation of Fig.~\ref{fig:YukawaWick2}(f) gives
the Hamiltonian term,
\beq
h_{t,2}^{(f)}
\es
g^2
\int[q_1 q_2 q_3]_\epsilon
\, \tdelta_{123}
\frac{\theta_\epsilon(q_1^+)}{q_1^+}
\frac{\theta_\epsilon(q_2^+)}{q_2^+}
\, B_{t,123.(-1)(-2)(-3)}
\, f_{t_r,123}^2
% \\
% \times
\mathrm{Tr}\left[
(\slashed q_1 - m)(\slashed q_2 + m)
\right]
\cN\!\left[
\Phi(q_3) \Phi(-q_3)
\right] ,
\label{eq:ht2f}
\eeq
where
\beq
\mathrm{Tr}\left[
(\slashed q_1 - m)(\slashed q_2 + m)
\right]
\es
2 \, \frac{(k^\perp)^2 + (2 x - 1)^2 m^2}{x(1-x)}
\ ,
\eeq
where $x = q_1^+/q_3^+$, and $k^\perp =
(1-x) q_1^\perp - x q_2^\perp$. The effective mass
of a boson depends on the longitudinal momentum of
that boson in a manner similar to the fermion case.

This concludes the calculation of the effective
Hamiltonian. Next, we renormalize the initial,
bare Hamiltonian by adding required counterterms
to render the effective Hamiltonian finite.

\subsection{Renormalized Yukawa Hamiltonian}
\label{sec:renoYukawa}

With the effective Hamiltonian terms evaluated in the preceding
sections, we can now determine whether the boundary condition
at $t = 0$ needs to be modified with counterterms, and if so,
what those counterterms need to be. For this purpose we need
to check whether matrix elements of the effective Hamiltonian
diverge when $t_r \to 0$. Matrix elements that can be represented
with diagrams with loops are most likely to produce divergences,
but in some cases also tree-diagram matrix elements can
diverge~\cite{Serafin:2023pkf}. Therefore, one needs to check
all matrix elements of $H_t$ between states that belong to the
domain of $H_t$.

From a mathematical point of view, the goal is to define $H_t$
that is a self-adjoint operator. By Stone's theorem, self-adjoint
operators are exactly those that generate strongly continuous
one-parameter groups, i.e., those and only those that can be
used to define unitary time evolution operators of the form
$e^{-\frac{i}{2} H_t x^+}$ in our case. It is common to first
define an operator that is symmetric, and then extend it into
a self-adjoint one. In practice, it is easiest to work with the
matrix elements of the Hamitonian, i.e., the quadratic form of
the Hamiltonian operator. Therefore, our main objective is to
show that, given appropriate counterterms, the limit as
$\epsilon^+ \to 0$ and $t_r \to 0$ of the quadratic form of
the effective Hamiltonian exists and is a symmetric quadratic
form.

For illustration purposes, we consider the kinetic Hamiltonian
restricted to the space of single-fermion states,
\beq
\ket{F}
\es
\sum_\sigma \int[p] \frac{\theta(p^+)}{p^+}
\psi_\sigma(p)
b_{p \sigma}^\dag \ket{0} ,
\label{eq:1particle}
\eeq
where $\psi_\sigma(p)$ is a single-particle wave function.
The $1/p^+$ factor is a convention that fits well with
the relativistic normalization of the operators,
$\{b_p, b_{p'}^\dagger\} = p^+ \tdelta(p-p')$. Whenever
two operators are contracted they produce a $p^+$ that is
canceled by a $1/p^+$ appearing in integrals such as in
Eq.~(\ref{eq:1particle}). Analogously, we define state
$\ket{F'}$ with wave function $\psi'_\sigma(p)$.

A single-fermion state $\ket{F}$ belongs to the Fock
space if and only if
\beq
\braket{F}{F}
\es
\sum_\sigma \int[p] \frac{\theta(p^+)}{p^+}
\left|\psi_\sigma(p)\right|^2
\label{eq:norm1}
\eeq
is finite. $\ket{F}$ belongs to the domain of the kinetic
Hamiltonian if $H_\text{kinetic} \ket{F}$ belongs to
the Fock space, i.e., if
\beq
\left\lVert H_\text{kinetic} \ket{F} \right\rVert^2
\es
\sum_\sigma \int[p] \frac{\theta(p^+)}{p^+}
\left| \frac{m^2 + (p^\perp)^2}{p^+} \psi_\sigma(p) \right|^2
\label{eq:domainHkin}
\eeq
is finite. If $\psi_\sigma(p) = m \sqrt{\cP^+ p^+} (\cP^+
+ p^+)^{-1} [m^2 + (p^\perp)^2]^{-1}$, where $\cP^+$ is
some parameter of longitudinal momentum dimension, then
$\braket{F}{F}$ is finite, but $\left\lVert H_\text{kinetic}
\ket{F} \right\rVert$ is infinite. Therefore, $\ket{F}$
with this wave function does not belong to the domain of
the kinetic Hamiltonian. On the other hand, functions of
the form $(p^+)^{\alpha + k} e^{-p^+/\cP^+} (p^1)^l (p^2)^n
e^{-(p^\perp)^2/m^2}$, for $\alpha > 1$ and $k, l, n \in
\mathbb{N}$ all belong to the domain of $H_\text{kinetic}$.
The set of finite linear combinations of these functions
is a dense subspace of the single-fermion Fock
sector~\cite{szego1975}. One then shows that $H_\text{kinetic}$
with the above-defined domain is self-adjoint in the Hilbert
space equal to the single-fermion Fock sector. The argument
can be extended to the full Fock space.

The above sketch of the proof of self-adjointness can in
principle be extended to the effective Hamiltonian $H_t$.
In doing so, however, one encounters a difficulty -- $H_t$
needs to be squared, which is rather tedious because $H_t$
is a complicated operator. One can avoid squaring $H_t$ if,
instead, one chooses to study the quadratic form of the
operator, i.e., the matrix elements of $H_t$. For the purpose
of this article this should suffice, since we do not rigorously
address the question of self-adjointness of $H_t$. For an
interested reader we offer a few remarks and leave the details
for future work.

Matrix elements of the kinetic Hamiltonian between
single-fermion states are,
\beq
\bra{F} H_\text{kinetic} \ket{F'}
\es
\sum_\sigma \int[p] \frac{\theta(p^+)}{p^+}
\frac{m^2 + (p^\perp)^2}{p^+}
\psi^*_\sigma(p)
\psi'_\sigma(p)
\ .
\label{eq:matElKin}
\eeq
It is not very difficult to see that if wave functions
$\psi$ and $\psi'$ are in the domain of $H_\text{kinetic}$,
then they lead to a well-defined matrix element, compare
Eqs.~(\ref{eq:domainHkin}) and (\ref{eq:matElKin}).

Questions about domains of operators and their forms
revolve around the behavior of the wave functions in
the neighborhoods of $p^+ = 0$, $p^+ = \infty$, and
$p^\perp = \infty$. Momenta belong to open sets,
$p^+ \in ]0, \infty[$, $p^\perp \in \mathbb{R}^2$.
For simplicity we consider wave functions with compact
support. This means that the integrands that define
matrix elements have no singularity around $p^+ = 0$,
and no slow decay when $p^+$ or $p^\perp \to \infty$,
which otherwise could lead to a divergent integral.
At the same time, compactly supported functions are
dense in the space of functions that define elements
of the Fock space. An additional complication is that
the Fock space consists of infinitely many Fock sectors,
each with a fixed number of particles. Hence, even if
in every Fock sector the wave function is compactly
supported, summing contributions from infinitely many
Fock sectors can still be divergent. Therefore, we
would like to define $H_t$ as an operator whose domain
is the subspace of the Fock space whose elements have
nonzero contributions in finitely many Fock sectors,
and the wave functions associated with each Fock sector
are compactly supported. So-defined domain is dense in
the Fock space. It follows rather easily that $H_t$ is
symmetric on its domain, meaning $\bra{\psi} H_t \ket{\phi}
= \bra{\phi} H_t \ket{\psi}^*$ for every $\ket{\psi}$
and $\ket{\phi}$ in the domain of $H_t$. However, there
is an obstacle: one first has to show that $H_t$ is
an operator. In other words one has to show that
$H_t \ket{\psi}$ belongs to the Fock space for every
$\ket{\psi}$ in the domain of $H_t$. To show it one
has to, effectively, compute the square of the Hamiltonian,
which, again, requires considerable amount of work.
Therefore, to avoid these complications we define $H_t$
to be the quadratic form with the same domain as described
previously, and given by the matrix elements of the
formal expressions for $H_t$ described in Secs.~\ref{sec:effHam}
through \ref{sec:H2self}. Since $H_t$ is narrow and
consequently better behaved than the bare Hamiltonian,
we expect that establishing a symmetric operator $H_t$
as well as a rigorous proof of existence of self-adjoint
extensions of this operator should be achievable.
Nevertheless, we do not attempt it here. Instead,
we assume that at least one such extension exists
as long as all matrix elements of $H_t$ are well-defined.

Interesting questions might be raised at this point.
Does the restriction to compactly supported wave functions
constitute a regularization similar to $t_r$ or $\epsilon^+$,
and are counterterms needed to remove it? The restriction
is different, since it does not have a cutoff and one can
always take a wave function whose support extends slightly
further toward infinity or slightly closer to zero. However,
bound states are expected to have wave functions that are
not compactly supported. Moreover, they are expected to have
nonzero contributions from infinitely many Fock sectors.
Therefore, bound states can only be realized as limits
of sequences of compactly supported Fock states that
approximate the targeted state better and better. Such
limits do not belong to the domain of the quadratic form
or the symmetric operator that we described earlier,
but they do belong to the domain of the self-adjoint
operator (if it exists) that extends the symmetric one.
Thus, the difficulty of proving that the above-mentioned
limits are well-defined is part of the difficulty of
proving that the self-adjoint extensions exist. Possibly
the simplest way to do it is to show that the quadratic
form $H_t$ is bounded from below. If it corresponds to
a symmetric operator, then its boundedness guarantees
that at least one self-adjoint extension of the symmetric
operator, the Friedrichs extension, exists~\cite{Reed:1975uy}.
Therefore, even though at this moment we are unable to
answer the above questions definitively, there is a clear
path forward that is likely to succeed in proving that
compactly supported wave functions do not introduce
a new kind of regularization that would require introduction
of counterterms. The question that we intend to answer
now is whether the limit $\epsilon^+, t_r \to 0$ exists
for every matrix element, $\bra{\psi} H_t \ket{\phi}$.
Positive answer will establish existence of $H_t$ as
a quadratic form, which is a prerequisite for finding
self-adjoint Hamiltonian operators.

We start by checking matrix elements of the effective
Hamiltonian in the simplest nontrivial case -- between
two single-fermion states. In the second order, the
matrix element has only contributions from the fermion
self-energy terms,
\beq
\bra{F'} \cH_{t,2} \ket{F}
\es
\sum_\sigma
\int\frac{dq^+ d^2q^\perp}{16\pi^3 q^+} \theta_\epsilon(q^+)
\, \frac{\delta m^2}{q^+}
\psi'^*_\sigma(q)
\psi_\sigma(q)
\ .
\label{eq:matEl1}
\eeq
From now on we consider $\epsilon^+ = 0$. The mass can
be written as $\delta m^2 = I_F(t+t_r) - I_F(t_r)$, where
\beq
I_F(t)
\es
g^2
\int\frac{dx d^2 k^\perp}{16\pi^3 x (1-x)}
\, \frac{f_{t,23.1}^2}{\cM_{23}^2 - m^2}
\, \frac{(k^\perp)^2 + m^2 (1 + x)^2}{x}
\ .
\label{eq:I_F}
\eeq
The integral in Eq.~(\ref{eq:matEl1}) is convergent but
the matrix element is divergent when $t_r \to 0$ because
$\delta m^2$ is divergent in that limit. At $t = 0$ the form
factor, $f_{t,23.1}^2 = 1$. In that case Eq.~(\ref{eq:I_F})
is divergent because the integrand approaches $1/x$ as
$|k^\perp| \to \infty$. For any $t > 0$ the region of
large $|k^\perp|$ is regulated by the form factor, hence,
the integral is convergent, because there are no other
sources of divergences. One might think that integration
in the neighborhood of $x = 0$ or $x = 1$ may produce
divergences, because both $x$ and $1-x$ appear in the
denominator. This, however, is not the case. To show
this one can first write the integral in polar coordinates
$k = |k^\perp|$ and $\varphi$, then integrate over $\varphi$,
and change integration variable from $k$ to $\cM^2 =
(m^2 + k^2)/x + (\mu^2 + k^2)/(1-x)$. After the change
of variables the integrand no longer contains any $x$
or $1-x$ factors in the denominator. The divergence
can be summarized as
\begin{multline}
I_F(t_r)
=
\frac{g^2}{16\pi^2}
\Bigg\{
  \frac{1}{4} q_1^+ \sqrt{\frac{\pi}{2 t_r}}
+ \left(m^{2} + \mu^{2}\right)
  \left[
    \log{\left( \frac{\sqrt{2 m^4 t_r}}{q_1^+} \right)}
  + \frac{\gamma}{2}
  \right]
+ \frac{m^{2}}{2} - \mu^{2}
+ \frac{\mu^{4}}{m^{2}} \log{\left(\frac{\mu}{m} \right)}
\\
+ \frac{2 \mu^{3} \sqrt{4 m^{2} - \mu^{2}} }{m^{2}}
  \operatorname{atan}{\left(\frac{\sqrt{4 m^{2} - \mu^{2}}}{2 m + \mu} \right)}
\Bigg\}
+ o(1) \ ,
\quad (t_r \to 0)
\label{eq:IFdiv}
\end{multline}
where $\gamma\approx 0.577$ is the Euler-Mascheroni
constant, and $o(1)$ stands for all the terms that
vanish when $t_r \to 0$.

To remove the divergence we modify the initial condition
for the Hamiltonian flow by adding a mass counterterm.
The mass counterterm needs to diverge with $t_r \to 0$
in order to remove the divergence from the effective
self-interaction term, but is otherwise not uniquely
determined at this point. We define the mass counterterm,
\beq
\delta m_X^2
\es
I_F(t_r) + \delta m_\text{finite}^2
\ ,
\label{eq:mXfermion}
\eeq
where $I_F(t_r)$ is the divergent part of the counterterm
and $\delta m_\text{finite}^2$ is the finite part.
$\delta m_\text{finite}^2$ is added to the initial
Hamiltonian, therefore, cannot depend on $t$. Also, it
cannot depend on $t_r$ in a divergent way. Otherwise
$\delta m_\text{finite}^2$ is at this point unspecified.

Our next goal is to fix the finite part of the mass
counterterm. In order to achieve this goal, we assume
that the physical mass squared of the fermion particle
is fixed at $m^2$ up to second order in the coupling
constant $g$. The physical mass means the value of
mass one can infer from the Hamiltonian eigenvalue
corresponding to an eigenstate describing a single
fermion. Such an eigenstate is called a physical fermion.
The physical mass is not equal to the bare mass that
would stand in a renormalized Lagrangian (typically
denoted $m_0$ in the literature), nor is it equal to the
mass parameter that stands in the effective Hamiltonian,
cf. $m_t$ in Eq.~(\ref{eq:sumMass}). The condition
we describe is known as the on-shell renormalization
condition.

In order to calculate the physical mass of a fermion up
to second order in $g$, one can use perturbation formulas
for the energy shift in the second order. To calculate
it correctly, one needs to take into account that the
perturbation (interaction Hamiltonian) contains terms
of order both $g$ and $g^2$. As always, first we specify
the unperturbed state. We choose $b_{P \lambda}^\dag \ket{0}$,
which is the same as the previously defined $\ket{F}$ with
$\psi_\sigma(p_1) = p_1^+\tdelta(p_1 - P) \delta_{\lambda\sigma}$.
The unperturbed eigenvalue of the effective Hamiltonian for
this state is $E^{(0)} = [m^2 + (P^\perp)^2]/P^+$. The
first-order correction is zero, because there are no terms
of order $g$ that have nonzero matrix elements between
single-fermion states. In the second order, the formula
requires energy denominators and one needs to evaluate
matrix elements of the first-order Hamiltonian terms
between the unperturbed state and states with one fermion
and one boson. The resulting correction is $-I_F(t+t_r)/P^+$.
Another second-order contribution comes from the
expectation value in the unperturbed state of the
second-order self-interaction Hamiltonian term. This
contribution is $[\delta m^2 + \delta m_X^2]/P^+
= [\delta m_\text{finite}^2 + I_F(t+t_r)]/P^+$.
Therefore, the total second-order correction is $E^{(2)}
= \delta m_\text{finite}^2/P^+$. In order to have
physical mass equal $m$, the second-order energy
correction needs to vanish. Hence,
\beq
\delta m_\text{finite}^2 \es 0 \ .
\eeq
With this form of the finite part of the counterterm,
the counterterm itself can be written in the following
form,
\beq
h_{\text{fermion mass }X,2}
\es
g^2
\int[q_1 q_2 q_3]_\epsilon
\, \tdelta_{23.1}
\, \frac{ f_{t_r,23.1}^2 }{ q_2^+  q_3^+ (q_2^- + q_3^- - q_1^-) }
\,\cN\!\left[
\bar\Psi(q_1)
(\slashed q_2 + m)
\Psi(q_1)
\right]
\nt
\left[
  \theta_\epsilon(q_2^+) \theta_\epsilon(q_3^+)
- \theta_\epsilon(-q_2^+) \theta_\epsilon(-q_3^+)
\right]
\\
\es
\int[q]_\epsilon \frac{\delta m_X^2}{q^+}
\,\cN\!\left[
    \bar\Psi(q)
    \frac{\gamma^+}{2}
    \Psi(q)
  \right] .
\label{eq:YukawaMassXF}
\eeq
The inclusion of the fermion mass counterterm in the
initial Hamiltonian implies inclusion of a new Wick's
diagram, Fig.~\ref{fig:YukawaWick2}(h).

In a completely analogous way we study the single-boson
matrix elements. The state of interest is now
\beq
\ket{B}
\es
\int[p_1] \frac{\theta(p_1^+)}{p_1^+}
\psi(p_1)
a_{p_1}^\dag \ket{0}
\ .
\eeq
We calculate the matrix element,
\beq
\bra{B'} \cH_{t,2} \ket{B}
\es
\int\frac{dq^+ d^2q^\perp}{16\pi^3 q^+}
\, \frac{\delta \mu^2}{q^+}
\psi'^*(q)
\psi(q)
\ ,
\label{eq:matEl2}
\eeq
where
\beq
\delta\mu^2
\es
g^2
\int[q_1 q_2]
\, q^+ \tdelta_{12.q}
\frac{\theta(q_1^+)}{q_1^+}
\frac{\theta(q_2^+)}{q_2^+}
\, \frac{f_{t+t_r,12.q}^2 - f_{t_r,12.q}^2}{(q_1 + q_2)^2 - \mu^2}
\, \mathrm{Tr}\left[
(\slashed q_1 - m)(\slashed q_2 + m)
\right] .
\label{eq:YukawaDeltaMu2}
\eeq
The matrix element in Eq.~(\ref{eq:matEl2}) is divergent
when $t_r \to 0$ because $\delta \mu^2$ is divergent in
that limit. For $4 m^2 > \mu^2$, the mass can be written
as $\delta \mu^2 = I_B(t+t_r) - I_B(t_r)$, where
\beq
I_B(t)
\es
2 g^2
\int\frac{dx d^2 k^\perp}{16\pi^3 x (1-x)}
\, \frac{f_{t,12.q}^2}{\cM_{12}^2 - \mu^2}
\, \frac{(k^\perp)^2 + (2 x - 1)^2 m^2}{x(1-x)}
\ ,
\eeq
where $x = q_1^+/q_3^+$, and $k^\perp = (1-x) q_1^\perp
- x q_2^\perp$. The condition on masses, $4 m^2 > \mu^2$,
ensures that the denominator $\cM_{12}^2 - \mu^2$ can never
be zero. For $4 m^2 < \mu^2$ the problem is not with RGPEP
evolution, $\delta\mu^2$ is still well-defined in that
case (even though $I_B$ is not well-defined without some
prescription for the singularity), but the scalar can
decay into a fermion-antifermion pair which leads to
technical complications when solving the theory. With
the assumption $2 m > \mu$, the same kind of analysis
as for the physical fermion state leads to the counterterm
$\delta\mu_X^2 = I_B(t_r) + \delta\mu_\text{finite}^2$,
and to the conclusion that $\delta\mu_\text{finite}^2 = 0$.
The integral $I_B$ is,
\begin{multline}
I_B(t_r)
=
\frac{g^2}{8\pi^2}
\Bigg\{
  \frac{1}{2} q^+ \sqrt{\frac{\pi}{2 t_r}}
+ \left(6 m^{2} - \mu^{2}\right)
  \left[\log{\left(\frac{3 m^2 \sqrt{2 m^4 t_r}}{(4 m^{2} - \mu^{2})q^+} \right)} + \frac{\gamma}{2}\right]
- m^{2}
\\
+ \frac{2 \left(4 m^{2} - \mu^{2}\right)^{\frac{3}{2}} }{\mu}
  \operatorname{atan}{\left(\frac{\mu}{\sqrt{4 m^{2} - \mu^{2}}} \right)}
\Bigg\}
+ o(1)
\ .
\quad(t_r \to 0)
\end{multline}
Therefore, the counterterm is,
\beq
h_{\text{scalar mass }X,2}
\es
g^2
\int[q_1 q_2 q_3]_\epsilon
\, \tdelta_{123}
\frac{\theta_\epsilon(q_1^+)}{q_1^+}
\frac{\theta_\epsilon(q_2^+)}{q_2^+}
\, \frac{f_{t_r,123}^2}{q_1^- + q_2^- + q_3^-}
% \nt
\mathrm{Tr}\left[
(\slashed q_1 - m)(\slashed q_2 + m) \right]
\cN\!\left[ \Phi(q_3) \Phi(-q_3) \right]
\\
\es
\frac{1}{2}
\int[q]_\epsilon \, \delta\mu_X^2
\,\cN\!\left[ \Phi(-q) \Phi(q) \right] .
\label{eq:YukawaMassXB}
\eeq
The counterterm is represented by a Wick diagram,
Fig.~\ref{fig:YukawaWick2}(i).

\begin{figure}
\includegraphics{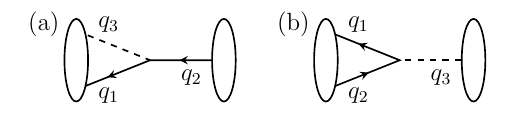}
\caption{\label{fig:YukawaMatEls1st}Two examples
of matrix elements of the first-order interaction
Hamiltonian $\cH_{t,1}$ evaluated between states
of different particle content. The two examples
correspond to Eqs.~(\ref{eq:1stFB_F}) and
(\ref{eq:1stFFbar_B}), respectively. The middle part
depicts the Wick's diagram, Fig.~\ref{fig:YukawaWick1}.
The ovals on the left and on the right represent
wave functions in the corresponding Fock sectors.
One can imagine that ovals come with a set of lines
representing the particle content, and to obtain
the matrix element one has to contract all lines.
For simplicity the dots that represent operators
or contractions are not drawn.}
\end{figure}

Now, we study matrix elements of the first-order interaction
Hamiltonian $\cH_{t,1}$. Two types of such matrix elements
are depicted in Fig.~\ref{fig:YukawaMatEls1st}. Matrix
element \ref{fig:YukawaMatEls1st}(a) requires a fermion-boson
state,
\beq
\ket{FB}
\es
\sum_{\sigma_1}
\int[p_1 p_3] \frac{\theta(p_1^+)}{p_1^+}
\frac{\theta(p_3^+)}{p_3^+}
\psi_{\sigma_1}(p_1, p_3)
b_{p_1, \sigma_1}^\dag a_{p_3}^\dag \ket{0} ,
\eeq
while matrix element \ref{fig:YukawaMatEls1st}(b) requires
fermion-antifermion state,
\beq
\ket{F \bar F}
\es
\sum_{\sigma_1, \sigma_2}
\int[p_1 p_2]
\frac{\theta(p_1^+)}{p_1^+}
\frac{\theta(p_2^+)}{p_2^+}
\psi_{\sigma_1,\sigma_2}(p_1, p_2)
b_{p_1,\sigma_1}^\dag d_{p_2,\sigma_2}^\dag \ket{0} .
\eeq
Therefore, the matrix elements,
Fig.~\ref{fig:YukawaMatEls1st}(a) and
Fig.~\ref{fig:YukawaMatEls1st}(b) are,
\beq
\bra{FB} \cH_{t,1} \ket{F}
\es
g
\sum_{\sigma_1, \sigma_2}
\int[q_1 q_2 q_3]
\frac{\theta(-q_1^+)}{|q_1^+|}
\frac{\theta( q_2^+)}{|q_2^+|}
\frac{\theta(-q_3^+)}{|q_3^+|}
\, \tdelta_{123}
\, f_{t + t_r,123}
% \nt
\bar u_{\sigma_1}(-q_1) u_{\sigma_2}(q_2)
\, \psi^*_{\sigma_1}(-q_1, -q_3) \psi_{\sigma_2}(q_2)
\, ,
\label{eq:1stFB_F}
\\
\bra{F \bar F} \cH_{t,1} \ket{B}
\es
g
\sum_{\sigma_1, \sigma_2}
\int[q_1 q_2 q_3]
\frac{\theta(-q_1^+)}{|q_1^+|}
\frac{\theta(-q_2^+)}{|q_2^+|}
\frac{\theta( q_3^+)}{|q_3^+|}
\, \tdelta_{123}
\, f_{t + t_r,123}
% \nt
\bar u_{\sigma_1}(-q_1) v_{\sigma_2}(-q_2)
\, \psi^*_{\sigma_1,\sigma_2}(-q_1, -q_2) \psi(q_3)
\ ,
\nn
\label{eq:1stFFbar_B}
\eeq
respectively. The integrands factor into functions coming
from the Hamiltonian and the wave functions. The former
are continuous while the latter are compactly supported.
Therefore, the integrals are always well-defined. One can
safely take the $t_r \to 0$ limit. Other types of matrix
elements, not depicted in Fig.~\ref{fig:YukawaMatEls1st},
include $\bra{F} \cH_{t,1} \ket{F B}$, $\bra{\bar F B}
\cH_{t,1} \ket{\bar F}$, etc. In all cases expressions
for matrix elements look similar to Eqs.~(\ref{eq:1stFB_F})
and (\ref{eq:1stFFbar_B}) (it is easy to guess what they
are!), and are well-defined.

\begin{figure}
\includegraphics{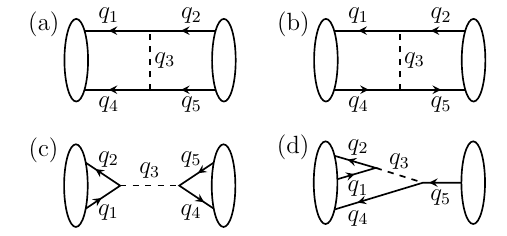}
\caption{\label{fig:YukawaMatElsA}Four examples of
matrix elements of the interaction Hamiltonian term
$\cH^{(a)}_{t,2}$. See Fig.~\ref{fig:YukawaMatEls1st}
for explanation.}
\end{figure}

Next, we turn our attention to more complicated matrix
elements. First, we study matrix elements of the interaction
Hamiltonian term $\cH^{(a)}_{t,2} = e^{-t(i\pd_f^-)^2}
h^{(a)}_{t,2}$. There are several types of matrix
elements one needs to consider, see
Fig.~\ref{fig:YukawaMatElsA}. The first one we analyze is
obtained by sandwiching $\cH^{(a)}_{t,2}$ between states,
\beq
\ket{F F}
\es
\sum_{\sigma_1, \sigma_2}
\int[p_1 p_2]
\frac{\theta(p_1^+)}{p_1^+}
\frac{\theta(p_2^+)}{p_2^+}
\,\psi_{\sigma_1, \sigma_2}(p_1, p_2)\,
b_{p_1\sigma_1}^\dagger
b_{p_2\sigma_2}^\dagger
\ket{0} .
\eeq
and analogously defined $\ket{F F'}$, where $\psi$
is replaced with $\psi'$. This is the example in
Fig.~\ref{fig:YukawaMatElsA}(a),
\beq
\bra{FF} \cH_{t,2}^{(a)} \ket{FF'}
\es
- g^2
\sum_{\sigma_1, \sigma_2, \sigma_4, \sigma_5}
\int[q_1 q_2 q_3 q_4 q_5]_\epsilon
\frac{\theta_\epsilon(-q_1^+)}{|q_1^+|}
\frac{\theta_\epsilon( q_2^+)}{|q_2^+|}
\frac{\theta_\epsilon(-q_4^+)}{|q_4^+|}
\frac{\theta_\epsilon( q_5^+)}{|q_5^+|}
\frac{\theta_\epsilon(q_3^+)}{q_3^+}
\nt
\, \tdelta_{1425}
\, \tdelta_{45.3}
\, f_{t,1425}
\, B_{t,123.45(-3)}
\, \bar u_{\sigma_1}(-q_1) u_{\sigma_2}(q_2)
\, \bar u_{\sigma_4}(-q_4) u_{\sigma_5}(q_5)
\nt
\left[
  \psi^*_{\sigma_1,\sigma_4}(-q_1, -q_4)
- \psi^*_{\sigma_4,\sigma_1}(-q_4, -q_1)
\right]
\left[
  \psi'_{\sigma_5, \sigma_2}(q_5, q_2)
- \psi'_{\sigma_2, \sigma_5}(q_2, q_5)
\right] .
% \nn
\label{eq:YukawaMatElA1}
\eeq
We also provide an explicit form of the matrix element
in Fig.~\ref{fig:YukawaMatElsA}(d). We define,
\beq
\ket{F F \bar F}
\es
\sum_{\sigma_1, \sigma_2, \sigma_3}
\int[p_1 p_2 p_3]
\frac{\theta(p_1^+)}{p_1^+}
\frac{\theta(p_2^+)}{p_2^+}
\frac{\theta(p_3^+)}{p_3^+}
\,\psi_{\sigma_1, \sigma_2, \sigma_3}(p_1, p_2, p_3)\,
b_{p_1\sigma_1}^\dagger
b_{p_2\sigma_2}^\dagger
d_{p_3\sigma_3}^\dagger
\ket{0} .
% \nn
\eeq
Hence, the matrix element is,
\beq
\bra{F F \bar F} \cH_{t,2}^{(a)} \ket{F}
\es
g^2
\sum_{\sigma_1, \sigma_2, \sigma_4, \sigma_5}
\int[q_1 q_2 q_3 q_4 q_5]_\epsilon
\, \frac{\theta_\epsilon(-q_1^+)}{|q_1^+|}
\, \frac{\theta_\epsilon(-q_2^+)}{|q_2^+|}
\, \frac{\theta_\epsilon(-q_4^+)}{|q_4^+|}
\, \frac{\theta_\epsilon( q_5^+)}{|q_5^+|}
\, \frac{\theta_\epsilon(q_3^+)}{q_3^+}
\nt
\, \tdelta_{1245}
\, \tdelta_{45.3}
\, f_{t,1425}
\, B_{t,123.45(-3)}
\, \bar u_{\sigma_1}(-q_1) v_{\sigma_2}(-q_2)
\, \bar u_{\sigma_4}(-q_4) u_{\sigma_5}(q_5)
\nt
\Big[
  \psi^*_{\sigma_4, \sigma_1, \sigma_2}(-q_4, -q_1, -q_2)
- \psi^*_{\sigma_1, \sigma_4, \sigma_2}(-q_1, -q_4, -q_2)
\Big]
\psi_{\sigma_5}(q_5)
\ .
\label{eq:YukawaMatElA2}
\eeq
Just as in the case of the first-order Hamiltonian,
the main difference between the two matrix elements
resides in the wave functions and how they are contracted
with the interaction kernel, which in both cases is very
similar. In fact, one can make the interaction kernel
in Eqs.~(\ref{eq:YukawaMatElA1}) and (\ref{eq:YukawaMatElA2})
identical by utilizing the relation between fermion
spinor $u$ and antifermion spinor $v$. However, the
difference between different types of matrix elements
of $\cH^{(a)}_{t,2}$ is irrelevant for the analysis
of convergence.

Compactly supported wave functions ensure convergence
of the integrals in the endpoint regions of $q_i^+ = 0,
\infty$ and $|q_i^\perp| = \infty$ for $i = 1, 2, 4, 5$.
Therefore, one needs to check whether the interaction
kernel contains any singularities away from the endpoints.
Spinors are singular only at the endpoints, hence,
do not obstruct convergence of the integrals. The
obvious candidate for a singularity is the $q_3^+ = 0$
point. Wave functions cannot regulate $1/q_3^+$ because
$q_3^+$ can become zero when all $q_1^+$, $q_4^+$,
$q_2^+$, and $q_5^+$ are nonzero and finite. However,
the singularity is only apparent, because, using momentum
conservation,
\beq
\frac{f_{t,1245} B_{t,123.45(-3)}}{q_3^+}
\es
\frac{1}{2}
\left(
  \frac{1}{(q_1 + q_2)^2 - \mu^2}
+ \frac{1}{(q_4 + q_5)^2 - \mu^2}
\right)
\left( f_{t,1245} - f_{t,123} f_{t,45.3} \right) .
% \nn
\eeq
The above expression appears in all versions of the
matrix elements involving $\cH_{t,2}^{(a)}$.
When particle 2 is annihilated while particle 1 is created,
as in Eq.~(\ref{eq:YukawaMatElA1}), $(q_1 + q_2)^2 \le 0$.
Hence, $(q_1 + q_2)^2 - \mu^2$ never vanishes in this
case and does not produce any singularities in the integrand.
More generally, whenever one among the particles 1 and 2,
or 4 and 5, is annihilated and the other is created, then
the corresponding denominator is strictly negative and
does not produce any singularities. When both 1 and 2
are created, as in Eq.~(\ref{eq:YukawaMatElA2}), or
annihilated, then $(q_1 + q_2)^2$ is the invariant
mass of the pair, hence, greater than or equal to $4 m^2$.
If $4 m^2 > \mu^2$, then the denominator, $(q_1 + q_2)^2
- \mu^2$ is strictly positive and does not lead to any
singularities of the integrand. If $4 m^2 \le \mu^2$,
then the denominator can vanish. However,
\beq
f_{t,1245} - f_{t,123} f_{t,45.3}
\es
\exp\left(
- t \frac{ \left[ (q_1 + q_2)^2 - (q_4 + q_5)^2 \right]^2 }{ (q_3^+)^2 }
\right)
\nm
\exp\left(
- t \frac{ \left[ (q_1 + q_2)^2 - \mu^2 \right]^2
         + \left[ (q_4 + q_5)^2 - \mu^2 \right]^2 }{ (q_3^+)^2 }
\right) .
\label{eq:formfactorsA}
\eeq
Therefore, whenever $(q_1 + q_2)^2 - \mu^2$ or $(q_4 + q_5)^2
- \mu^2$ vanishes, while $q_3^+$ stays nonzero, $f_{t,1245}
- f_{t,123} f_{t,45.3}$ vanishes as well in a way that makes
$f_{t,1245} B_{t,123.45(-3)} / q_3^+$ finite.

The analysis can become much more complicated when both
$(q_1 + q_2)^2 - \mu^2$, and $q_3^+$ vanish at the same
time. This can happen only in examples (c) and (d) in
Fig.~\ref{fig:YukawaMatElsA} and only when $4 m^2 \le \mu^2$.
However, in those cases $q_3^+ = 0$ implies $q_1^+ = q_2^+
= 0$, which is excluded by the compactness of the wave
functions. We could stop the analysis here, but it is
worthwhile to check what would happen if we removed the
restriction on the wave functions. It turns out that
Fig.~\ref{fig:YukawaMatElsA}(d), is strongly regulated
by the exponents in Eq.~(\ref{eq:formfactorsA}) -- whenever
$q_3^+ \to 0$, the form factors $f_{t,1245} - f_{t,123}
f_{t,45.3}$ approach zero exponentially quickly.
Figure~\ref{fig:YukawaMatElsA}(c) on the other hand
is much more complicated. $f_{t,1245} - f_{t,123} f_{t,45.3}$
might stay nonzero when the limit $q_3^+ \to 0$ is taken
along certain directions in the multidimensional space
of arguments of the integrand. Such highly singular
behavior is difficult to analyze.

\begin{figure}
\includegraphics{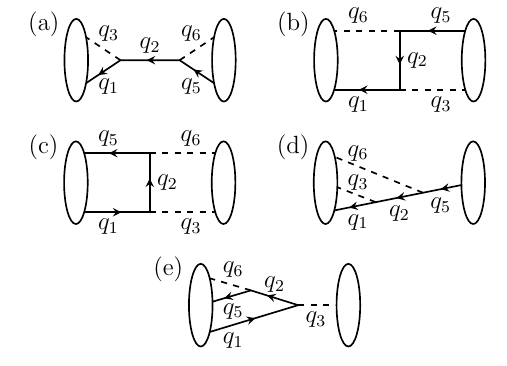}
\caption{\label{fig:YukawaMatElsBC}Five types of matrix
elements of the interaction Hamiltonian term
$\cH^{(b)}_{t,2} + \cH^{(c)}_{t,2}$. See
Fig.~\ref{fig:YukawaMatEls1st} for explanation.}
\end{figure}

There are two additional terms that need to be analyzed.
One of them is represented in Fig.~\ref{fig:YukawaMatElsBC},
which shows five examples of matrix elements of
$\cH^{(b)}_{t,2} + \cH^{(c)}_{t,2}$, see Eq.(\ref{eq:ht2bc}).
The last term is $\cH^{(0)}_{t,2} = e^{-t(i\pd_f^-)^2}
H_{\psi^2\phi^2}$, see Eq.~(\ref{eq:Hpsi2phi2reg}). Matrix
elements of $\cH^{(0)}_{t,2}$ can be represented by the same
diagrams as in Fig.~\ref{fig:YukawaMatElsBC}. Therefore, it is
most convenient to analyze both terms together. We have,
\begin{multline}
  \cH_{t,2}^{(0)}
+ \cH_{t,2}^{(b)}
+ \cH_{t,2}^{(c)}
=
g^2
\int[q_1 q_2 q_3 q_5 q_6]_\epsilon
\, \theta_\epsilon(|q_2^+|)
\, f_{t,1356}
\, \tdelta_{1356}
\, \tdelta_{56.2}
f_{t_r,123}
f_{t_r,56.2}
\\
\times
\cN\!\left\{ \wick{
\bar\Psi(-q_1) \Phi(q_3)
\left[
  A \frac{ \gamma^+ }{ 2 }
+ \frac{ B_{t,123.(-2)56} }{ q_2^+ }
  \left( \frac{1}{2} q_2^+ \gamma^- - q_2^\perp \gamma^\perp + m \right)
\right]
\Phi(q_6) \Psi(q_5)
} \right\} ,
\label{eq:cHt0bc}
\end{multline}
where
\beq
A
\es
  \frac{ 1 }{ q_2^+ }
+ \frac{ B_{t,123.(-2)56} }{ q_2^+ }
  \, \frac{m^2 + (q_2^\perp)^2}{q_2^+}
\label{eq:A1}
\\
\es
  \frac{1}{2}
  \left(
    \frac{-q_1^- - q_3^-}{(q_1 + q_3)^2 - m^2}
  + \frac{q_5^- + q_6^-}{(q_5 + q_6)^2 - m^2}
  \right)
\nm
  \left(
    \frac{1}{(q_1 + q_3)^2 - m^2}
  + \frac{1}{(q_5 + q_6)^2 - m^2}
  \right)
  \frac{m^2 + (q_2^\perp)^2}{2 q_2^+}
\frac{f_{t,123} f_{t,56.2}}{f_{t,1356}}
\ .
\label{eq:A2}
\eeq
This form of the expression is most suitable for analysis
and is motivated by the fact that $\slashed q_2$ contains
potentially singular $1/q_2^+$. 

The explicit $1/q_2^+$ in Eq.~(\ref{eq:cHt0bc}) does not
lead to any singularities because
\beq
\frac{ B_{t,123.(-2)56} }{ q_2^+ }
\es
\frac{1}{2}
\left(
  \frac{1}{(q_1 + q_3)^2 - m^2}
+ \frac{1}{(q_5 + q_6)^2 - m^2}
\right)
\left(
1 - \frac{f_{t,123} f_{t,56.2}}{f_{t,1356}}
\right) .
\eeq
Momenta $q_1$ and $q_3$ correspond to a fermion and a boson,
respectively. Depending on the signs of $q_1^+$ and $q_3^+$
the respective particle is either created or annihilated.
When both particles are created or both particles are
annihilated, $(q_1 + q_3)^2 \ge (m + \mu)^2$, and the
denominator $(q_1 + q_3)^2 - m^2$ is always strictly greater
than zero. When one is created and one is annihilated,
$(q_1 + q_3)^2 \le (m - \mu)^2$. Importantly, $q_2^+ \to 0$,
implies $q_1^+ = -q_3^+$, meaning that $q_1^-$ and $q_3^-$
have different signs. Therefore, $q_1^- + q_3^-$ stays
finite when $q_2^+ \to 0$ and we have $(q_1 + q_3)^2 =
-(q_1^\perp + q_3^\perp)^2 \le 0$ in the limit. Hence, the
denominator $(q_1 + q_3)^2 - m^2$ cannot vanish when $q_2^+
\to 0$. It can vanish only when $q_2^+$ is nonzero and
$4 m^2 \le \mu^2$. In that case however, the form factors
$1 - f_{t,123} f_{t,56.2}/ f_{t,1356}$ vanish at the same
time, making the expression well-defined. For particles
5 and 6, one can repeat above arguments to show no
singularities. For the same reason $A$ in Eq.~(\ref{eq:cHt0bc})
is also regular when either $(q_1 + q_3)^2 - m^2$ or
$(q_5 + q_6)^2 - m^2$ vanishes, see Eq.~(\ref{eq:A1}).
Finally, when $q_2^+ \to 0$ only Fig.~\ref{fig:YukawaMatElsBC}(b)
and Fig.~\ref{fig:YukawaMatElsBC}(c) need to be analyzed
-- in other cases wave functions regulate the integrals.
The first term in Eq.~(\ref{eq:A2}) is not a source of
any problems because the denominators cannot vanish when
$q_2^+ \to 0$. The second term in Eq,~(\ref{eq:A2}) also
leads to no singularities because when $q_2^+ \to 0$,
we have $(q_1 + q_3)^2 \le 0$ and $(q_5 + q_6)^2 \le 0$.
Therefore,
\beq
f_{t,123} f_{t,56.2}
\es
e^{-t\frac{[(q_1 + q_3)^2 - m^2]^2}{(q_2^+)^2}}
e^{-t\frac{[(q_5 + q_6)^2 - m^2]^2}{(q_2^+)^2}}
\eeq
vanishes exponentially quickly.

We conclude the analysis of the matrix elements. Every
term in the effective Hamiltonian has been analyzed
using the simplest relevant matrix elements. There
are infinitely many other matrix elements that contain
spectators -- particles that do not contract with the
interaction Hamiltonian. For the purpose of finding
divergences it suffices to check only the matrix elements
that do not contain spectators. Only mass terms required
introduction of counterterms at second order. At higher
orders one expects further contributions to the mass
counterterms as well as coupling constant counterterms.

\subsection{\label{sec:summary}Summary
of the renormalized Hamiltonian}

We summarize all terms of the renormalized effective
Hamiltonian in the second order of perturbative
calculations. Inclusion of the counterterms allows
us to remove all regularization factors.

The kinetic energy terms together with the mass
counterterms are,
\beq
\cH_\text{mass}
\es
\int[q] \frac{m_t^2(q) + (q^\perp)^2}{q^+}
\,\cN\!\left[
    \bar\Psi(q)
    \frac{\gamma^+}{2}
    \Psi(q)
  \right]
+
\int[q] \, \frac{\mu_t^2(q) + (q^\perp)^2}{2}
\,\cN\!\left[ \Phi(-q) \Phi(q) \right] ,
\label{eq:sumMass}
\eeq
where the effective mass of the fermion is
\beq
m_t^2(q_1)
\es
m^2
+
g^2
\int[q_2 q_3]
\theta(q_2^+) \theta(q_3^+)
\frac{ q_1^+ \tdelta_{23.1} }{ q_2^+ q_3^+ }
\, \frac{f_{t,23.1}^2}{(q_2 + q_3)^2 - m^2}
\, (2 q_1 \cdot q_2 + 2 m^2)
\ ,
\eeq
while the effective mass of the boson is,
\beq
\mu_t^2(q_3)
\es
\mu^2
+
g^2
\int[q_1 q_2]
\theta(q_1^+)
\theta(q_2^+)
\, \frac{ q_3^+ \tdelta_{12.3} }{q_1^+ q_2^+}
\, \frac{f_{t,12.3}^2}{(q_1 + q_2)^2 - \mu^2}
\, \mathrm{Tr}\left[
(\slashed q_1 - m)(\slashed q_2 + m)
\right] .
% \nn
\eeq
The effective fermion mass squared is the sum of the kinetic
mass squared $m^2$ and second-order self-interaction contributions
represented in diagrams Fig.~\ref{fig:YukawaWick2}(d),
Fig.~\ref{fig:YukawaWick2}(e), and Fig.~\ref{fig:YukawaWick2}(h).
The effective boson mass squared is the sum of the kinetic
mass squared $\mu^2$ and second-order self-interaction
contributions represented in diagrams
Fig.~\ref{fig:YukawaWick2}(f) and Fig.~\ref{fig:YukawaWick2}(i).

The first-order effective interaction term, depicted in
Fig.~\ref{fig:YukawaWick1}, is,
\beq
\cH_{t,1}
\es
g
\int[q_1 q_2 q_3]
\, \tdelta_{123}
\, f_{t,123}
\, \cN\!\left[ \bar\Psi(-q_1) \Psi(q_2) \Phi(q_3) \right] .
\eeq

There are two second-order interaction terms,
effective four-fermion interaction,
\beq
\cH_{t,2}^{(\Psi^4)}
\es
\cH_{t,2}^{(a)}
\rs
g^2
\int[q_1 q_2 q_3 q_4 q_5]
\, \tdelta_{1245}
\, \tdelta_{45.3}
\, f_{t,1245}
\, \frac{B_{t,123.45(-3)}}{q_3^+}
\, \theta(q_3^+)
% \\
% \times
\cN\!\left[
\bar\Psi(-q_1) \Psi(q_2)
\bar\Psi(-q_4) \Psi(q_5)
\right] ,
\label{eq:HtPsi4}
\eeq
which is represented with Fig.~\ref{fig:YukawaWick2}(a),
and the effective instantaneous fermion interaction,
\beq
\cH_{t,2}^{(\Psi^2\Phi^2)}
\es
g^2
\int[q_1 q_2 q_3 q_5 q_6]
\, \tdelta_{1356}
\, \tdelta_{56.2}
\, f_{t,1356}
% \\
% \times
\,
\cN\!\left\{ \wick{
\bar\Psi(-q_1) \Phi(q_3)
\left[
  \frac{ \gamma^+ }{ 2 q_2^+ }
+ \frac{ B_{t,123.(-2)56} }{ q_2^+ }
  ( \slashed{q}_2 + m )
\right]
\Phi(q_6) \Psi(q_5)
} \right\} .
\nn
\eeq
The effective instantaneous fermion interaction
is composed of contributions from diagrams in
Fig.~\ref{fig:YukawaWick2inst},
Fig.~\ref{fig:YukawaWick2}(b),
and Fig.~\ref{fig:YukawaWick2}(c).
An equivalent form of Eq.~(\ref{eq:HtPsi4}) is one
with $\theta(q_3^+)$ replaced with $1/2$.

\section{Conclusion}
\label{sec:conclusion}

In this article we provide explicit formulas for the
effective renormalized Hamiltonian of Yukawa theory
calculated up to terms order $g^2$ using the renormalization
group procedure for effective particles. The result is
summarized in Sec.~\ref{sec:summary}. The Hamiltonian
can be utilized in numerical calculations within
frameworks of Discretized Light Cone Quantization,
Basis Light Front Quantization, or any other suitable
numerical approach. To our knowledge this is the most
extensive RGPEP analysis performed to date. It is also
a necessary starting point for quantum simulations
on future quantum computers.

The main advantage of using an effective Hamiltonian
instead of the bare one is that ultraviolet divergences,
typical to quantum field theories, are absent. One
implication is that once free parameters of the theory
are fixed, there is no need to change them whenever
the truncated basis within a numerical calculation is
enlarged. This is not the case with the bare Hamiltonian
for which mass and coupling constant need to be modified
depending on the size of the basis in order to obtain
finite results for observables. Therefore, $H_t$
allows for variational calculations (in the subspace
of fixed longitudinal momentum $P^+$).
Furthermore, any uncertainties inherent to the choice
of the numerical approach used are independent of
the uncertainties stemming from the approximation
of dropping all terms order $g^3$ or higher. One can
therefore, study them separately in a controlled
manner.

In order to perform the calculations efficiently we
introduced Wick's diagrams. This development is helpful
because it allows us to handle many types of interaction
terms all at once. Table~\ref{tab:3list} illustrates this
point well -- all eight types of terms are contained
in a simple expression of Eq.~(\ref{eq:Hpsi2phi})
and represented with a single diagram,
Fig.~\ref{fig:YukawaWick1}. The reduction in the number
of terms is especially important
in higher-order calculations -- one additional leg
in a diagram corresponds roughly to the doubling
of the number of terms written using creation and
annihilation operators.

The Wick's diagrams we developed are defined in momentum
space, hence, differ from those in position space considered
previously~\cite{Wick:1950ee,Coleman:2018mew}. Since we
study effective Hamiltonians, instead of the scattering
matrix, we define contractions that are not symmetric with
respect to transposition of contracted operators. Because
of this, the diagrams we use cannot be deformed freely -- one
cannot change the initial order of vertices. On the other hand,
because RGPEP Eq.~(\ref{eq:RGPEP}) contains commutators,
only connected diagrams contribute to the effective Hamiltonians.
It might be possible to introduce a different kind of
diagrams, one in which order of vertices does not matter,
and the symmetries of diagrams are fully used. Such
diagrams would simplify higher-order calculations
further, but are not needed in the current, second-order
calculation.

Presented calculations are formal from a mathematically
rigorous point of view. They can be made rigorous if one
interprets them as describing quadratic forms rather than
operators. We showed that after the addition of counterterms
the cutoffs can be removed and the resulting symmetric quadratic
form, $H_t$ is well-defined on the domain of finite linear
combinations of states that have finitely many nonzero
Fock components and each component has compactly supported
wave functions. An interesting continuation of this line of
research is to first prove that this form corresponds to
a symmetric operator, and then to determine whether this
symmetric operator is essentially self-adjoint or if it admits
multiple self-adjoint extensions. By showing that the form
is bounded from below one can prove that at least one self-adjoint
extension exists, the Friedrichs extension~\cite{Reed:1975uy}.
The possibility that multiple self-adjoint extensions exist is
exciting, because if true, it is likely to distinguish different
physics associated to the small-$p^+$ region of the Fock space.
In other words, zero-mode physics, associated with the vacuum,
might be differentiated by different self-adjoint extensions
of $H_t$. In this context one might ask questions such as,
what self-adjoint extension is approached when one tries to
recover the continuum limit of a discretized theory? Do basis
function frameworks approach the same self-adjoint extension?
What does one need to do to approach a different self-adjoint
extension? Some vacuum effects may need to be included via
special counterterms that seem not to be easily discoverable
using cutoff Hamiltonians alone, an example is given in
Appendix A in Ref.~\cite{Wilson:1994fk}. These special
counterterms would introduce additional terms in the effective
Hamiltonians. Can those special counterterms be found by
studying self-adjoint extensions of the symmetric Hamiltonian
operator? And so on. Finally, a rigorous formulation of the
unitary transformation utilized by RGPEP is also of interest.

\begin{acknowledgments}
This material is based upon work supported by the U.S.
Department of Energy, Office of Science, National Quantum
Information Science Research Centers, Quantum Systems
Accelerator.
Kamil Serafin and Peter Love acknowledge support from
U.S. Department of Energy through Grant DE-SC0023707 under
the Office of Nuclear Physics Quantum Horizons program
for the ``Nuclei and Hadrons with Quantum computers (NuHaQ)''
project.
Carter Gustin was supported by the Exclusives
via Artificial Intelligence and Machine Learning (EXCLAIM)
collaboration, DOE grant DE-SC0024644.
\end{acknowledgments}

\appendix
\section{Notation and conventions}
\label{sec:notation}

The summation symbol used in Sec.~\ref{sec:effectiveParticles}
is defined as follows.
\beq
\sum_i
\es
\sum_\text{discrete}
\int_0^\infty\frac{dp_i^+}{4\pi p_i^+}
\int\frac{d^2p_i^\perp}{(2\pi)^2}
\ ,
\eeq
where $\sum_\text{discrete}$ means summation over all discrete
quantum numbers of the particle characterized by $i$. Moreover,
the operator normalization is given by,
\beq
q(i) q(j)^\dagger \pm q(j)^\dagger q(i)
\es
\delta_\text{discrete}
p_i^+ \tdelta^3(p_i - p_j)
\ ,
\eeq
where, assuming $i$ and $j$ correspond to the same type of
particles, the plus sign is chosen for fermions, the minus
sign is chosen for bosons, $\delta_\text{discrete}$ stands
for Kronecker delta between corresponding discrete quantum
numbers of particle $i$ and particle $j$, and the momentum
conservation Dirac delta,
\beq
\tdelta^3(p)
\es
4\pi \delta(p^+) (2\pi)^2 \delta^2(p^\perp)
\ .
\eeq

The representation of the Dirac algebra we are using is
\begin{align}
\gamma^0
&\rs
\left[\begin{matrix}
0 & 1 \\
1 & 0
\end{matrix}\right]
 ,
\quad
\gamma^3
\rs
\left[\begin{matrix}
0 & -1 \\
1 & 0
\end{matrix}\right]
 ,
\\
\gamma^1
&\rs
\left[\begin{matrix}
-i\sigma^2 & 0 \\
0 & i\sigma^2
\end{matrix}\right]
 ,
\quad
\gamma^2
\rs
\left[\begin{matrix}
i\sigma^1 & 0 \\
0 & -i\sigma^1
\end{matrix}\right]
 ,
\end{align}
where each entry represents a two-by-two matrix, and
$\sigma^1$, and $\sigma^2$ are the Pauli matrices.
The spinors are,
\beq
u_\sigma(p)
\es
\frac{1}{\sqrt{p^+}}
\left[\begin{matrix}
p^+ \xi_\sigma \\
(-i \sigma^2 p^1 + i \sigma^1 p^2 + m)\xi_\sigma
\end{matrix}\right]
 ,
\\
v_\sigma(p)
\es
\frac{1}{\sqrt{p^+}}
\left[\begin{matrix}
-p^+ \xi_{-\sigma} \\
(i \sigma^2 p^1 - i \sigma^1 p^2 + m)\xi_{-\sigma}
\end{matrix}\right]
 ,
\eeq
where $\xi_\sigma = [ \delta_{\sigma,\frac{1}{2}},
\delta_{\sigma,-\frac{1}{2}} ]^T$ is a two-vector. The spinors
satisfy the Dirac equation,
\beq
(\slashed p - m) u_\sigma(p)
\rs
(\slashed p + m) v_\sigma(p)
\es
0
\ ,
\label{eq:spinorDirac}
\eeq
and additional relations,
\beq
\sum_\sigma u_\sigma(p) \bar u_\sigma(p) \es \slashed p + m \ ,
\\
\sum_\sigma v_\sigma(p) \bar v_\sigma(p) \es \slashed p - m \ .
\eeq
From Eqs.~(\ref{eq:spinorDirac}) follows
\beq
\left( \frac{1}{2}i\pd_f^- \gamma^+ + \frac{1}{2}i\pd^+ \gamma^-
- i\pd^1 \gamma^1 - i\pd^2 \gamma^2 - m \right) \psi(x) \es 0 \ .
\eeq
Note that the time derivative is $\pd_f^-$ instead of $\pd^-$.

Additionally, we present a useful identity relating front-form
energy and invariant mass. Suppose in an interaction vertex
there are $k$ particles created and $n-k$ particles annihilated.
Due to momentum conservation, $p_1^+ + \dots + p_k^+ = p_{k+1}^+
+ \dots + p_n^+$, $p_1^\perp + \dots + p_k^\perp = p_{k+1}^\perp
+ \dots + p_n^\perp$. An important type of equation follows,
\beq
(p_1^- + \dots + p_k^-) - (p_{k+1}^- + \dots + p_n^-)
\es
\frac{(p_1 + \dots + p_k)^2 - (p_{k+1} + \dots + p_n)^2}
{p_1^+ + \dots + p_k^+}
\ ,
\eeq
where $(p_1 + \dots + p_k)^2 = (p_1 + \dots + p_k)^\mu
(p_1 + \dots + p_k)_\mu$ is the invariant mass squared of
the set of particles $1$ through $k$. In other words,
one can trade energy differences for differences in
squares of invariant masses.

% \bibliograpGustin:hystyle{plain}
% \bibliographystyle{unsrt} %Seems to fix the citation ordering problem.
\bibliography{citations}

\end{document}